\documentclass[fleqn,usenatbib,useAMS]{mnras}

\usepackage{graphicx}	
\usepackage{amsmath}	
\usepackage{amssymb}	
\usepackage{multicol}   
\usepackage{bm}		    
\usepackage{pdflscape}	
\usepackage{xcolor}	    
\usepackage{subcaption}
\usepackage{caption}
\usepackage[T1]{fontenc}
\usepackage{newtxtext,newtxmath}
\usepackage{ae,aecompl}

\title[The Magnetic Anatomy of a White Dwarf Chromosphere]{The Complex Magnetic Anatomy of a White Dwarf Chromosphere}

\author[Iskandarli et al.]{Leyla Iskandarli,$^1$\thanks{E-mail: iskandarlileyla@gmail.com}
Jay Farihi,$^1$
Stefano Bagnulo,$^2$
John D.~Landstreet,$^{2,3}$
Akshay Robert,$^1$
J.~J.~Hermes,$^4$
\newauthor
Steven G.~Parsons,$^5$
Nikolay~Walters,$^1$
Nick Achilleos$^{1,6}$
\\
$^1$Department of Physics and Astronomy, University College London, London WC1E 6BT, UK\\
$^2$Armagh Observatory and Planetarium, Armagh BT61 9DG, Northern Ireland, UK\\
$^3$Department of Physics and Astronomy, University of Western Ontario, London, Ontario N6A 3K7, Canada\\
$^4$Department of Astronomy and Institute for Astrophysical Research, Boston University, Boston, MA 02215, USA\\
$^5$Department of Physics and Astronomy, University of Sheffield, Sheffield, S3 7RH, UK\\
$^6$The Centre for Planetary Sciences at UCL/Birkbeck, London WC1E 6BT, UK
}

\date{Accepted XXX.  Received YYY; in original form ZZZ}
\pubyear{2026}

\begin{document}
\maketitle

\begin{abstract}
This study presents phase-resolved spectroscopy and spectropolarimetry of the DAHe white dwarf LP\,705-64.  Here, two apparently discrete chromospheric regions are observed to alternate in visibility and magnetic polarity with rotation, with field strengths of 9.5 and 22\,MG based on the measured Zeeman splitting in H$\upalpha$. As in the class prototype GD\,356, emission line equivalent widths vary in antiphase with the broadband light curve. The 9.5\,MG field is visible for 60\,per cent of the stellar rotation and stays nearly constant, whereas the 22\,MG field increases modestly during the remaining phases. Notably, the time-series transitions between magnetic regions are abrupt and transit-like in morphology, requiring at least one magnetic region to be highly localised and significantly smaller than the areas inferred for the dark surface spots.  Both the intensity and the Stokes $V/I$ spectra indicate a field configuration significantly more complex than an offset dipole, but higher-cadence data are necessary to distinguish between a global field and localised magnetic regions. All data are consistent with magnetic, chromospheric, and dark surface regions that are co-located and thus physically related. These findings support a model of chromospheric emission that is intrinsic and localised, consistent with predictions of dynamo-driven electrical currents that heat the upper atmosphere near magnetic poles.
\end{abstract}




\section{Introduction}

The existence of strong magnetic fields in white dwarfs was hypothesised nearly 80 years ago, based on simple rotation in analogy with the Earth and Sun \citep{Blackett1947}, and the first detection occurred more than 20 years later \citep{Kemp1970}.  Since that discovery over half a century ago, the underlying cause of magnetism in white dwarfs remains a fundamental yet unanswered question in stellar evolution.  Detected field strengths span four orders of magnitude, from $10^{4}$ to $10^{8}$\,G, exceeding the solar surface field (1\,G) and sunspot fields.  The earliest field detections in white dwarfs were achieved through circular polarization signatures of magnetism \citep[e.g.][]{Angel1970, Angel1974}, and soon after via Zeeman-split Balmer lines, providing direct measurements of their surface magnetic fields \citep{Liebert1975}.

Within the volume-complete 20\,pc sample, approximately 22\,per cent of isolated white dwarfs have magnetic fields that are detectable using spectropolarimetry \citep{Bagnulo2021}.  This can be compared to around 8\,per cent of white dwarfs in magnitude-limited surveys, where visible Zeeman splitting in low- to moderate-resolution optical spectra is required \citep{Wickramasinghe2005}.  The presence of Zeeman splitting in H$\upalpha$ has become the most common means of identifying magnetic white dwarfs, especially following the advent of large spectroscopic surveys such as the Sloan Digital Sky Survey \citep[SDSS; e.g.][]{Schmidt2003} and the Dark Energy Spectroscopic Instrument \citep[DESI; e.g.][]{Kilic2026,Swan2026}.  Spectropolarimetry continues to play a key role in characterising the properties of individual stars and magnetic white dwarfs as a population, and in particular by overcoming previous survey biases and by detecting weak fields where Zeeman splitting is absent in low-resolution intensity spectra \citep{Kawka2007, Jordan2007, Landstreet2012, Bagnulo2019, Landstreet2019}.

A few distinct origins have been proposed for these fields: fossil fields amplified by flux conservation during post–main sequence contraction \citep[][]{Woltjer1964, Landstreet1967, Tout2004}, convective dynamos arising in stellar mergers \citep[][]{Regos1995, Tout2008, Nordhaus2011, Garcia-Berro2012} or following giant-planet engulfment \citep[][]{Siess1999, Farihi2011, Kissin2015}, and convective dynamos driven by core crystallization during cooling \citep[][]{Horn1968, Isern2017, Ginzburg2022}.  The observations support a picture where fields emerge at cooling ages of two or three Gyr for remnants with masses consistent with single-star evolution \citep{Bagnulo2021}, and are superficially compatible with either a primordial field that reaches the surface at late times \citep{Camisassa2024}, or a field re-forged via crystallization \citep{Blatman2024}.  For white dwarfs with masses above 1.1\,M$_\odot$, which are likely merger products, magnetism is commonly detected at cooling ages less than 1\,Gyr, implying the merging process immediately results in strong surface fields \citep{Bagnulo2022}.

Stars of spectral type DAHe are a subclass of isolated magnetic white dwarfs with Zeeman-split Balmer lines in emission.  For nearly four decades the enigmatic prototype GD\,356 was the only known example \citep{Greenstein1985}, but now roughly two dozen are known, mostly via large spectroscopic surveys \citep{Reding2020, Ganiscke2020, Manser2023}.  All members of this class exhibit kG- to MG-level fields that typically result in detectable splitting in low-resolution spectra, and, where possible to constrain, rapid rotation with periods of minutes to hours.  Importantly, it has been shown that DAHe stars cluster on the Hertzsprung–Russell (HR) diagram \citep{Ganiscke2020}, which, combined with additional empirical and theoretical considerations, has been used to demonstrate that the emission must result from an intrinsic heating mechanism \citep{Walters2021}.  These predictions and the clustering in mass and temperature have been confirmed in larger samples \citep{Manser2023,Reding2023}.  A magnetic field associated with $^{22}$Ne distillation has been shown to be capable of producing currents that drive surface heating that is consistent with the observed properties of DAHe stars \citep{Lanza2024}.

Prior to this work, only the prototype GD\,356 has been observed with full-phase coverage of its rotation, using time-series data including multi-band photometry, spectroscopy, and spectropolarimetry.  Its light curve has a sinusoidal shape, implying that the dark surface spot is always in view but never fully exposed.  Both spectropolarimetry and Zeeman splitting reveal a mildly varying field, consistent with a single magnetic pole in view at all phases. The mean magnetic field modulus, $\langle |B| \rangle$, and the mean line-of-sight field component, $\langle B_{\rm z} \rangle$, are 11.5 and 3\,MG, respectively. Notably, in GD\,356 the emission feature strength peaks near photometric minimum, suggesting a temperature-inverted, optically thin emitting layer above the dark photospheric region \citep{Walters2021}. All of these data support a picture of a spatially coincident region shared by the observed magnetic pole, surface spot, and chromosphere.

LP\,705-64 is the brightest of three DAHe white dwarfs known to exhibit two distinct troughs in their light curves \citep{Manser2023, Farihi2023}.  Based on {\em TESS} data, there is a clear periodogram peak at 36.3\,min \citep{Reding2023}, but ground-based time-series photometry reveals this to be the second harmonic of the true rotation period of 72.6\,min, with a double-peaked light curve exhibiting two unequal minima per cycle \citep{Farihi2023}.  Modelling indicates there are two antipodal dark spots with angular radii at least as large as $20\degr$, although the inclination of the rotation axis and the co-latitude of the spots remain degenerate \citep{Farihi2023}.  Spectroscopy covering part of the rotation cycle reveals two distinct Zeeman splitting patterns in H$\upalpha$ emission, corresponding to field strengths of 9.4 and 22.2\,MG \citep{Reding2023}, thus sharing basic similarities with G183-35, which exhibits two sets of Zeeman triplets seen in absorption as it rotates \citep{Kilic2019}. These results for LP\,705-64 indicate that both magnetic poles rotate into and out of view, but do not constrain any spatial or temporal correlations between the dark surface spots, magnetic field, and chromospheric regions of the star.

This paper presents a comprehensive analysis of LP\,705-64 based on phase-resolved spectroscopy and spectropolarimetry using X-shooter and the FOcal Reducer and low dispersion Spectrograph (FORS2).  The study quantifies the chromospheric Balmer emission line strengths, the magnetic field regions, and compares these with published light curve data.  Section~\ref{sec:2} describes the observations and data, Section~\ref{sec:3} details the measurement procedures and the results, Section~\ref{sec:4} discusses the implications for chromospheric and magnetic geometries for LP\,705-64 and the DAHe class, and Section~\ref{sec:5} summarises the main findings.

\begin{table}
\begin{center}
\caption{Published parameters of LP\,705-64.}
\label{tab:Tab1}

\begin{tabular}{lr}

\hline

Parameter                       &Value\\

\hline

Spectral type                   &DAHe\\
$G$ (mag)                       &$16.890\pm0.003$\\
Distance (pc)                   &$52.7\pm0.2$\\
$v_{\rm tan}$ (km\,s$^{-1}$)    &$52.9\pm0.3$\\
$T_{\rm eff}$ (K)               &$8440\pm200$\\
log\,[$g$\,(cm\,s$^{-2}$)]      &$8.34\pm0.05$\\
Mass (M$_\odot$)                &$0.81\pm0.04$\\
$P_{\rm rot}$ (min)             &$72.629\pm0.004$\\

\hline

\end{tabular}
\end{center}
{\em References:} Astrometry and photometry \citep{Gaia2023}; stellar parameters \citep{Reding2023}; rotation period \citep{Farihi2023}.
\end{table}

\section{Observations and data}
\label{sec:2}
\subsection{Time-series spectroscopy}

\begin{figure*}
\includegraphics[width=\textwidth]{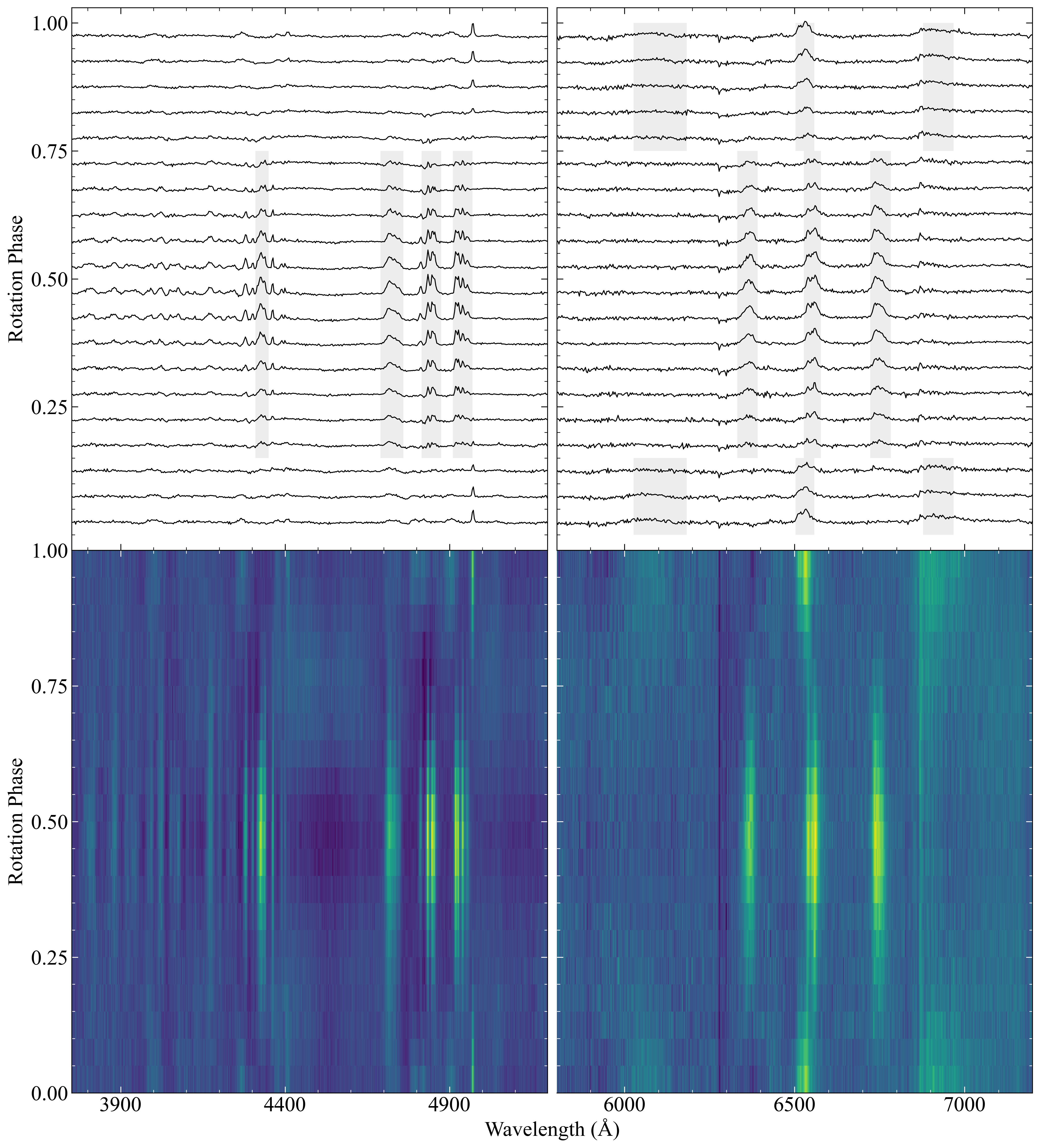}  
    \caption{\textit{Upper panels:} Rotation phase-folded X-shooter spectra, where phase $=0$ corresponds to the photometric minimum based on the published ephemeris \citep{Farihi2023}.  The 113 UVB and 118 VIS spectra were co-added into 20 bins of phase width 0.05 each, and spectrally binned to 3\,\AA\ for better visualisation. Telluric absorption residual flux in the range 6850--6950\,\AA\ is $\approx$\,11\,per cent for all VIS spectra. Grey shading marks the readily identifiable Balmer emission features at each phase; where Zeeman triplets can be discerned, the central feature is $\uppi$, with $\upsigma^-$ to the left and $\upsigma^+$ to the right. The spectra show two spatially distinct emission regions: narrow splitting that is apparent during phases 0.15--0.75, and broader splitting accompanied by significantly weaker $\upsigma^{\pm}$ features outside this range.  The dual structure is most evident in H$\upalpha$, where the $\uppi$ component variation can be clearly seen, whereas in H$\upbeta$ the broader Zeeman components cannot be unambiguously distinguished, and for H$\upgamma$ only the central emission component is clear during phases of narrow Zeeman splitting.  \textit{Lower panels:} The data from the upper panels plotted as trailed spectrograms.  The maximum relative flux is indicated in yellow and the minimum in blue.  Periodic variations in both the relative emission intensity and wavelength are visible in multiple features.  This figure can be compared and contrasted with the trailed H$\upalpha$ spectrogram of G183-35, which exhibits five Zeeman {\em absorption} components over a full rotation \citep{Kilic2019}.}
\label{fig:trailedspec}
\end{figure*}

Time-series spectra were obtained in visitor mode at the Very Large Telescope (VLT) at Paranal Observatory on 2022 October 28 and 29 using the triple-arm echelle spectrograph X-shooter \citep{Vernet2011}.  The instrument was used in its standard configuration, employing the UVB (3000--5595\,\AA) and VIS (5595--10\,240\,\AA) arms in staring mode, as the target is too faint to yield useful, phase-resolved data in the NIR (10\,240--24\,800\,\AA) instrument arm.  Slit widths of 1.3 and 1.2\,arcsec were used in the UVB and VIS arms, respectively, and in both cases low-gain readout with on-chip binning of 2 pixels in the spectral direction was employed to increase the signal-to-noise ratio (S/N).  The spectral resolving power achieved with this configuration was $R\approx4100$ in the UVB and $R\approx6500$ in the VIS arm.  

The X-shooter observations were conducted continuously, totalling approximately 6.3\,h on-source across the two nights, with individual integration times of 200\,s in the UVB arm and 192\,s in the VIS arm. The resulting cadence is close to 3.6\,min, yielding 113 and 118 frames in the UVB and VIS arms, respectively. This observing strategy ensured good phase coverage, with around 20 samples of each 72.63\,min rotation period, and several rotation cycles observed on each of the two nights.  The spectra were obtained under clear skies and typical Paranal conditions.  Standard calibration frames, including bias, flat-field, and arc lamp exposures, were obtained according to the standard X-shooter calibration plan, together with a single spectrophotometric standard star (HIP\,110573) for flux calibration.

Data reduction followed standard spectroscopic procedures using {\sc eso reflex} \citep{Goldoni2011, Freudling2013}.  Gaussian source localisation was applied to identify the target position along the slit, after which the UVB and VIS spectra of LP\,705-64 were traced using optimal extraction.  Cosmic rays were removed using a $3\upsigma$ clipping threshold.  The individual extracted spectra were continuum-normalized by fitting a second-order polynomial over spectral regions that were free of Balmer emission or telluric absorption features.  A correction was applied to the VIS spectra in the region of the telluric B band (6850--6950\,\AA) using the standard star as a reference.  This procedure reduced the telluric absorption in this region for all VIS spectra to an average residual of $\approx$\,11\,per\,cent in flux.  The S/N values assessed for individual spectra in the regions 4500--4700 and 6720--6800\,\AA\ are found to be $\langle{\rm S/N}\rangle = 34\pm2$ and $15\pm2$, respectively.

\subsection{Spectropolarimetry}

Spectropolarimetry was carried out on 2022 October 31 and November 1 with FORS2 \citep{Appenzeller1998} on the VLT.  As with the X-shooter data set, observations with FORS2 were conducted continuously over two full nights, covering several rotation cycles each night.  On the first night, the 1200R grating (5800--7300\,\AA) was used to observe H$\upalpha$, before switching to the 1200B grating (3700--5200\,\AA) for the second night to observe H$\upbeta$.  In both cases, a slit width of 1.0\,arcsec was used, achieving a nominal resolving power of $R\approx2100$ in the red and $R\approx1400$ in the blue.  The observing conditions were clear and typical for Paranal.

The observing strategy employed the retarder waveplate at alternating position angles of $\pm45\degr$ in an ABBA pattern.  This beam-swapping procedure removes systematic instrumental effects \citep[see e.g.][]{Bagnulo2009}.  The two spectra split by the Wollaston prism (ordinary and extraordinary beams) were co-added to form the Stokes $I$ intensity spectra, while the difference of the beam pairs yielded the circular polarization signal Stokes $V/I$.  Individual exposures had an integration time of 470--480\,s.  The spectropolarimetric observations provide simultaneous flux and circular polarization spectra, at cadences of around 8 and 16\,min, respectively.  

The data were pre-processed using standard {\sc iraf} routines for spectroscopy.  Beams split by the Wollaston prism and recorded at different retarder waveplate angles were recombined using standard beam-swapping formulae\footnote{See the FORS2 user manual \url{https://www.eso.org/sci/facilities/paranal/instruments/fors/doc.html}}.  The typical maximum S/N of the spectra obtained after recombining a pair of exposures was about 80 per \AA.  This strategy provided 24 individual intensity spectra and 12 Stokes $V$ profiles for each setup, offering a balance between phase resolution and S/N across the observing run.

\section{Analysis and results}
\label{sec:3}

The spectroscopy and spectropolarimetry time-series are analysed in the context of the previously published data.  The phase-folded light curves exhibit two unequal minima per cycle, consistent with two polar regions (Figure~\ref{fig:trailedspec} and ~\ref{fig:lcvsew}).  For simplicity and consistency, the hydrogen emission lines are referred to here as `narrow' and `broad', corresponding to the weaker and stronger of the two observed Zeeman splitting configurations, respectively.

\subsection{Emission line strength variation and wavelength shifts}
\label{sec:3.1}

The analysis carried out in this section is based on the X-shooter time-series spectroscopy.  A cursory assessment of the magnetic field variation may be obtained from the spectra as seen in Figure~\ref{fig:trailedspec}.  H$\upalpha$ is resolved into a Zeeman triplet, consisting of $\uppi$ (central), $\upsigma^{-}$ (blueshifted), and $\upsigma^{+}$ (redshifted) components.  The narrow triplet appears most clearly between phases 0.15 and 0.75, while the broad triplet appears outside these phases.  Visually, emission strength variation is most apparent in the narrow H$\upalpha$ $\upsigma^{\pm}$ lines where their amplitude peaks around phase 0.45.  The broad $\upsigma^{\pm}$ lines peak in strength near phase 1.  H$\upbeta$ likewise forms a Zeeman triplet, but the low amplitude and the blending of the broad components render them indistinct, with emission similarly apparent during phases 0.15--0.75.  The narrow triplet is resolved for the H$\upbeta$ components, and their emission strength variability peaks consistently with H$\upalpha$.  At H$\upgamma$, only a single emission feature is consistently detected over the phase range corresponding to the narrow features in  H$\upalpha$ and H$\upbeta$.

\begin{figure}
    \includegraphics[width=\columnwidth]{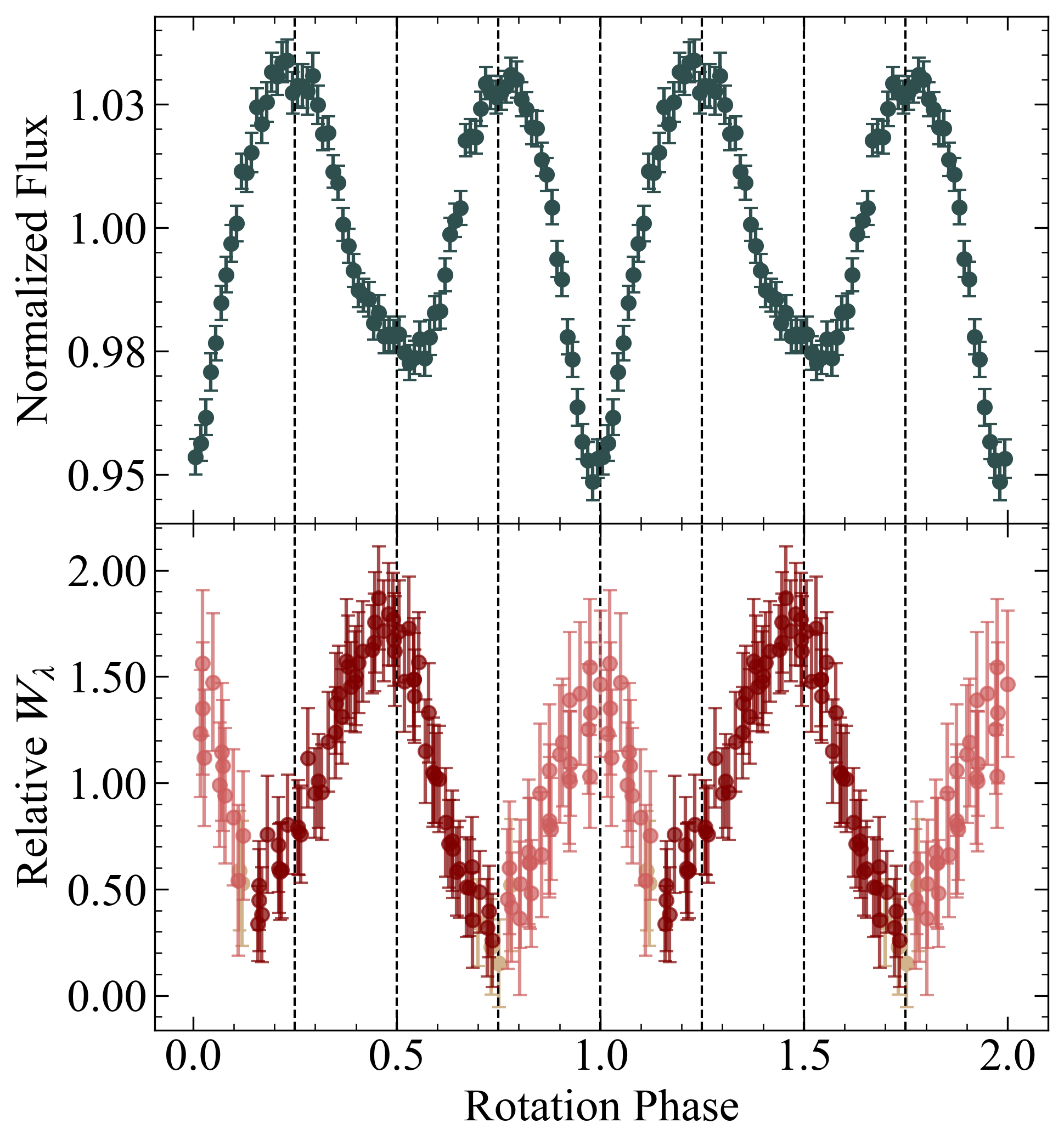}  
    \caption{Comparison of the X-shooter H$\upalpha$ emission line strengths to the \textit{g+r} band average light curve \citep{Farihi2023} of LP\,705-64 as a function of rotation phase.  The time-series photometry is shown in the upper panel, and the $\langle W_\uplambda \rangle$ emission time-series is shown in the lower panel.  Lighter red symbols denote $\langle W_\uplambda \rangle$ measurements of the broad emission triplet, while darker red symbols denote those of the narrow emission lines, which are discrete in wavelength and rotation phase. Measurements for the six (of 118) spectra that contain both narrow and broad emission features are shown in tan, where the observation timing has resulted in sampling of both magnetic regions in these single exposures.}
    \label{fig:lcvsew}
\end{figure}

Although the 3800--4100\,\AA\ region may appear noisy, the standard deviation in this spectral range demonstrates a clear correlation between phase and low-level emission peaks.  This is consistent with emission from the higher Balmer lines that is in phase with the narrow emission in H$\upalpha$ and H$\upbeta$, but decreasing in strength toward the higher energy levels of hydrogen.  Notably, there are no distinct emission or absorption features at \ion{Ca}{ii} H \& K, which sensitively trace external pollution in white dwarfs at this effective temperature \citep{Zuckerman2003,Koester2005}.

The equivalent widths\footnote{In contrast with convention and for simplicity, $W_{\uplambda}$ is treated as positive for the emission lines.}, $W_{\uplambda}$, of the Zeeman-split emission features are measured to quantify their variation in strength.  Because the H$\upbeta$ triplet components are blended during the phases when these features are broad, the $W_{\uplambda}$ analysis is confined to H$\upalpha$.  The individual component $W_{\uplambda}$ values are divided by the temporal mean, and all three are averaged into a single $\langle W_\uplambda \rangle$ per time-series exposure, which is then passed to the discrete Fourier analysis tool {\sc period04} \citep{Lenz2005}.

The periodogram for the $W_{\uplambda}$ time-series shows a peak signal at 39.64\,d$^{-1}$ (36.33\,min), which coincides with the peak frequency observed in a periodogram of the {\em TESS} light curve \citep{Reding2023}.  However, both of these are the second harmonic of the true photometric (and thus rotation) period, and result from the double-peaked nature of these time-series data sets \citep{Farihi2023}.  In the $W_{\uplambda}$ periodogram, there is a peak detected at the fundamental frequency of 19.83\,d$^{-1}$, corresponding to $72.61\pm0.03$\,min, and within the error of the established photometric period (Table~\ref{tab:Tab1}).  When $\langle W_\uplambda \rangle$ is phase folded on the photometric period in Figure~\ref{fig:lcvsew}, it displays a near-antiphase trend; with the exception of the modest offset near phase 0.25, the photometric maximum occurs when $\langle W_\uplambda \rangle$ is near minimum and vice versa. 

\begin{figure}
    \includegraphics[width=\columnwidth]{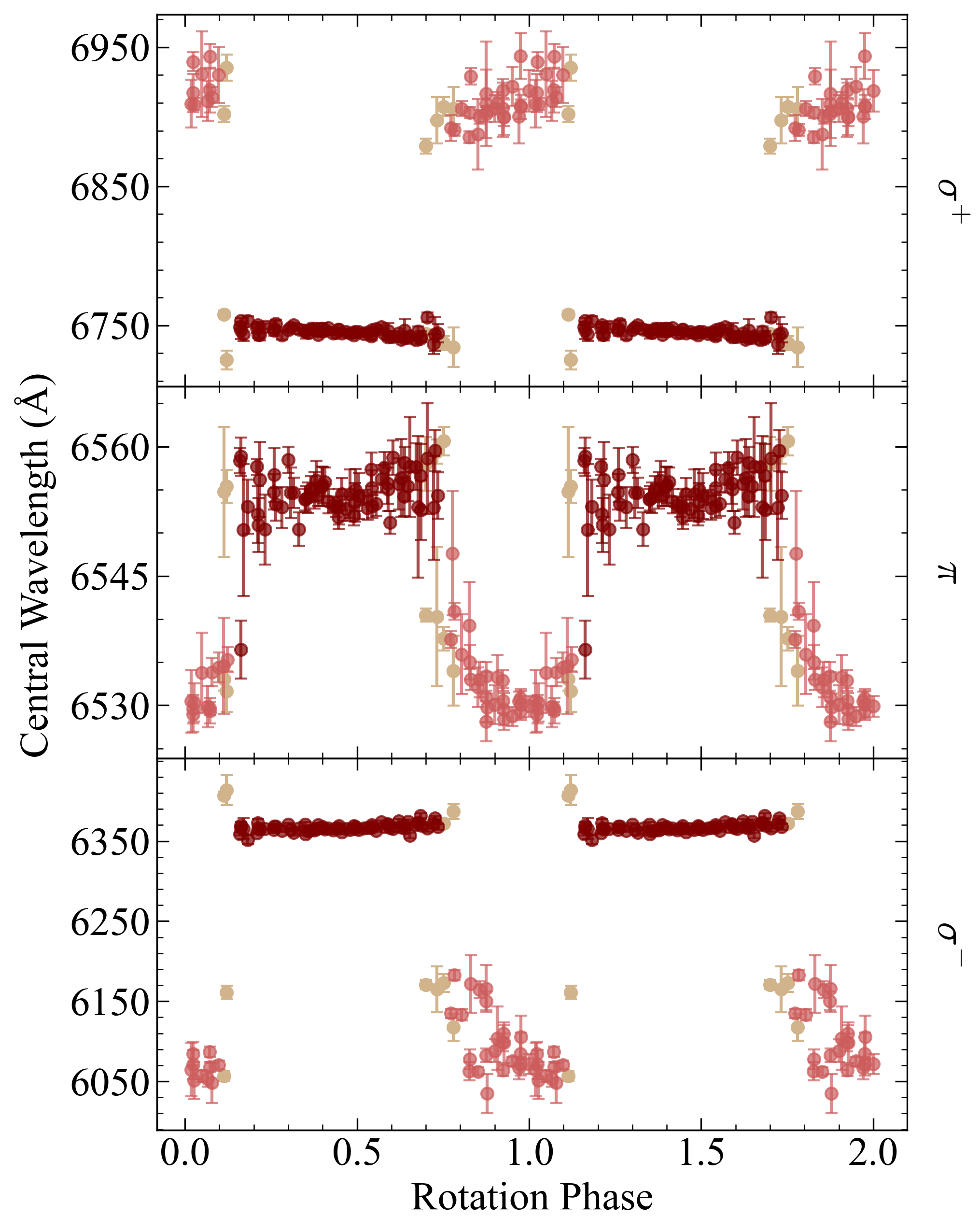}  
    \caption{Phase-folded central wavelengths of the H$\upalpha$ Zeeman triplet components in the X-shooter time-series, shown across two cycles for clarity, and labelled on the right side of the $y$-axis. The measurements are determined using Gaussian fits to the emission lines.  The data points are colour-coded identically to Figure~\ref{fig:lcvsew}, based on discrete sets of lines measured; dark red for the narrow splitting, light red for the broad splitting, and tan for the six (of 118) X-shooter spectra where both sets of emission lines are present owing to exposure length and timing.  There is no evidence for transitional smearing between the narrow and broad $\upsigma^{\pm}$ components in any spectra, including the six spectra where both sets are present.  However, the narrow and broad $\uppi$ components can be blended in these exposures with durations spanning the transition between the two magnetic regions (Section~\ref{sec:3.1}).}
    \label{fig:centralwavelengths}
\end{figure}

Each H$\upalpha$ emission component is fitted with a Gaussian profile, by using least-squares minimisation between the fit and the data, yielding the central wavelength.  These are plotted as a function of rotational phase in Figure~\ref{fig:centralwavelengths}.  These plots are informative, but care must be taken because of the limitations resulting from exposure time and cadence. Every spectrum shows the narrow or the broad emission features; only six of 118 show both, and none show anything in between.  That is, within these exposures of 192\,s, there are only the two discrete sets of $\upsigma^{\pm}$ pairs, and no indication of smearing that might be associated with a gradual shift from a hemisphere-averaged magnetic field of 9.5 to 22\,MG, or vice versa.  

Thus, owing to this discrete behaviour, each set of $\upsigma^{\pm}$ measurements is disconnected, reflecting the separation of the two sets in the actual spectra, both in wavelength and in rotation phase.  Nevertheless, the abrupt switch between narrow and broad Zeeman splitting could indicate two magnetically distinct emission regions that each rotate into and out of view once per spin period, at least one of which may be highly localised.  The sole exception is the $\uppi$ component of H$\upalpha$, which is always present at some level, with a median full width at half maximum of 40\,\AA.  Between 9.5 and 22\,MG, atomic data predict a wavelength shift around 50\,\AA\ for this $\uppi$ component, consistent with the spectra \citep[see Figure~\ref{fig:specpol};][]{Schimeczek2014}.

\subsection{Magnetic field determinations}
\label{sec:3.2}

\begin{figure*}
    \includegraphics[width=\textwidth]{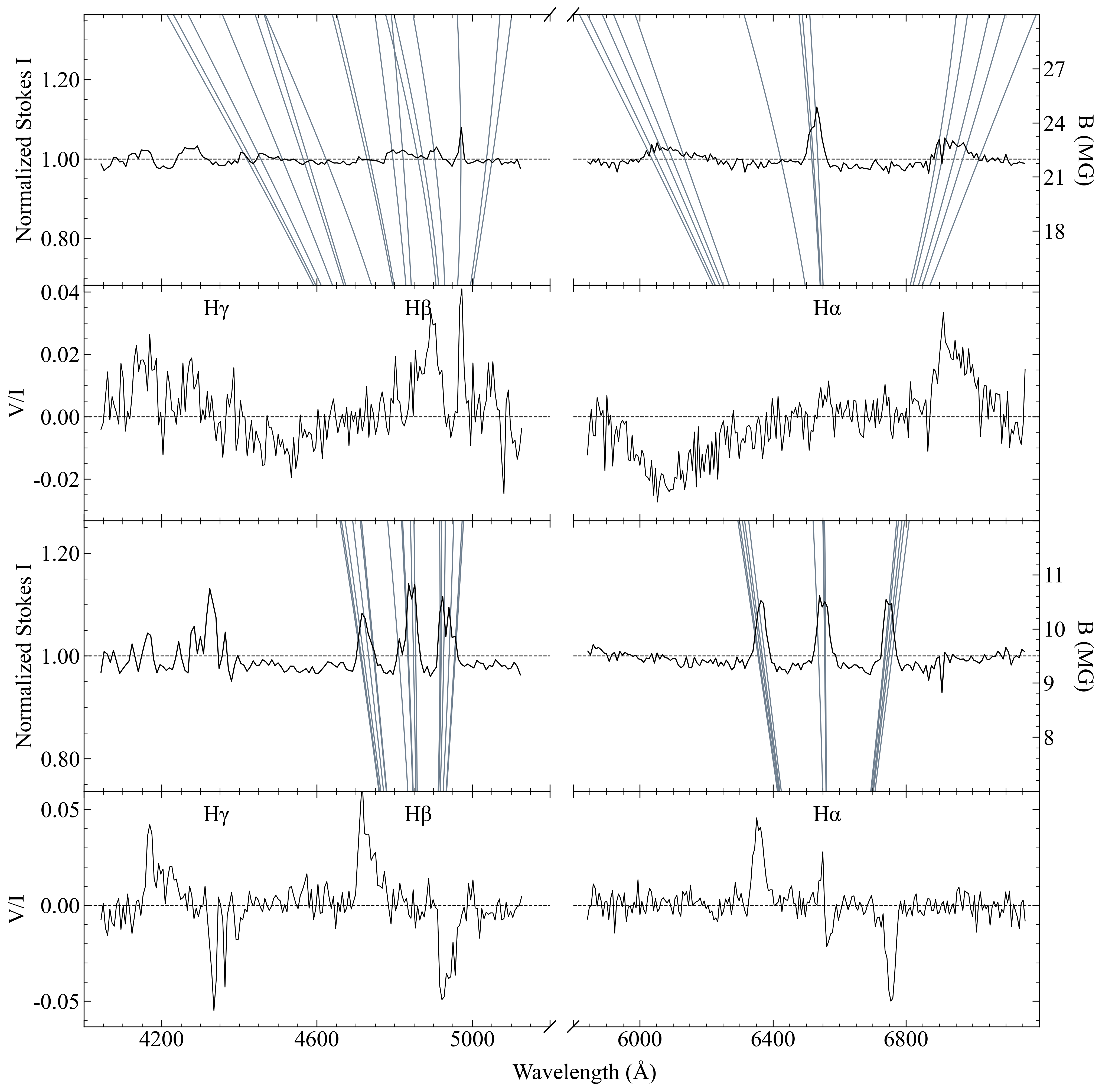}  
    \caption{FORS2 spectropolarimetry at two rotational phases.  The top two panels are at a rotational phase near 0.02 and depict the broad Zeeman-split features, and the bottom two panels are at a rotational phase close to 0.50 and depict the narrow emission lines.  From the top, the first and third panels show the normalized Stokes $I$ spectra in wavelength regions similar to Figure~\ref{fig:trailedspec}.  Grey curves overplot the theoretical wavelengths of the Zeeman-split components of H$\upalpha$, H$\upbeta$ and H$\upgamma$ as a function of magnetic field strength (right-hand axis), computed for fields of 22\,MG and 9.5\,MG \citep{Schimeczek2014}.  The second and fourth panels show the corresponding circular polarization spectra via Stokes $V/I$, where dashed horizontal lines mark the continuum levels.  All spectra have been continuum-normalized, corrected for telluric absorption, and binned to 8\,\AA.}
    \label{fig:specpol}
\end{figure*}

The spectropolarimetric data corroborate two magnetically distinct regions.  As in the non-polarimetric X-shooter data, the FORS2 spectra exhibit a narrow Zeeman triplet and a broader triplet that may rotate into and out of view.  The Stokes profiles plotted in Figure~\ref{fig:specpol} demonstrate that both Zeeman triplets have the characteristic dual polarity associated with the two circular polarization states, with opposite signs as expected for fields originating from opposite poles.  The interpretation of the spectropolarimetric data and Zeeman splitting follows standard practice \citep[e.g.][]{bagnulo2015}.

\begin{table}
\begin{center}
\caption{Phase-resolved magnetic field values from FORS2 spectropolarimetry.}
\label{tab:Bz}
\begin{tabular}{lrr}

\hline
Phase           &$\langle|B|\rangle$ (MG)   &$\langle B_{\rm z} \rangle$ (MG)\\
\hline  
0.14            &$9.41\pm0.01$              &$-4.75\pm0.52$\\
0.38            &$9.56\pm0.01$              &$-3.07\pm0.33$\\
0.62            &$9.39\pm0.02$              &$-4.91\pm0.50$\\
0.90            &$21.82\pm0.04$             &$8.86\pm0.41$\\
\hline
\end{tabular}
\end{center}
\end{table}

The surface-averaged, longitudinal magnetic field $\langle B_{\rm z} \rangle$ is measured as follows using least-squares fitting \citep[e.g.][]{Bagnulo2002}:

\begin{equation}
V_\uplambda = -g_{\rm eff} \, C_{\rm z} \, \uplambda^2 \frac{{\rm d}I_\uplambda}{{\rm d}\uplambda} \langle B_{\rm z} \rangle
\label{eqn1}
\end{equation}

\noindent\\
where $V_\uplambda$ and $I_\uplambda$ are the Stokes $V$ and $I$ profiles, respectively.  The effective Land\'e factor, $g_{\rm eff}$, is set to 1.0.  The constant is defined as $C_{\rm z}=e/4 \uppi m_ec^2 \approx 4.7\,\times\,10^{-13}$\,\AA$^{-1}$\,G$^{-1}$, where $e$ is the electron charge, $m_{\rm e}$ the electron mass, and $c$ the speed of light \citep{Landstreet1975}.

$\langle B_{\rm z} \rangle$ is measured from the 12 FORS2 Stokes $V/I$ spectra obtained with the 1200R grating; because $\langle|B|\rangle$ and $\langle B_{\rm z} \rangle$ are both derived from H$\upalpha$, the 1200B data are used only for the H$\upbeta$ and H$\upgamma$ panels of Figure~\ref{fig:specpol}. Because each $V/I$ measurement spans over 20\,per cent of a rotation cycle, and the timing of the observations naturally results in four clusters of three exposures each in phase, the 12 polarimetric spectra are combined into phase bins centred at 0.14, 0.38, 0.62, and 0.90.  $\langle B_{\rm z} \rangle$ is derived using the linear Zeeman approximation. At 22\,MG the individual H$\upalpha$ components are no longer displaced linearly with field strength, and the approximation is therefore inadequate for modelling the positions and widths of individual components (Section~\ref{sec:3.3}). The first moment of the Stokes $V$ profile is far less sensitive to this departure, however, and the approximation introduces only a small error in $\langle B_{\rm z} \rangle$ at the field strengths measured here. 

The resulting magnetic moments are listed in Table~\ref{tab:Bz}.  The phase during which the 22\,MG feature is in view yields $\langle B_{\rm z} \rangle \approx +8.9$\,MG, while the three phases containing the 9.5\,MG feature yield negative values ranging from $\langle B_{\rm z} \rangle \approx-3.1$\,MG  to $\langle B_{\rm z} \rangle \approx-4.9$\,MG.  The polarity reversal between the two regions is in direct agreement with the sign change of the Stokes $V/I$ profiles seen in Figure~\ref{fig:specpol}, and confirms that the narrow and broad sets of emission lines originate in magnetic regions of opposite sign.  Quoted uncertainties are formal fitting errors; realistic uncertainties are dominated by the instrument response correction of the intensity spectra, and are estimated to be a few per cent for $\langle|B|\rangle$, and at least 10--15\,per cent for $\langle B_{\rm z} \rangle$.

$\langle|B|\rangle$ is estimated from the Zeeman wavelength splitting of H$\upalpha$ using the classical relation \citep[e.g.][]{Landstreet1992}:

\begin{equation}
     \upDelta\uplambda = 4.67 \times 10^{-13} \, g_{\rm eff} \, \uplambda_0^2 \, \langle |B| \rangle
    \label{eq:zeeman}
\end{equation}

\noindent\\
where $\Delta\uplambda$ is the measured wavelength shift of each $\upsigma^{\pm}$ component from the line centre in \AA, $\uplambda_0$ is the rest wavelength of H$\upalpha$ in \AA, and $\langle|B|\rangle$ is the mean field modulus in G. For each exposure, flux-weighted centroids are measured for the $\upsigma^{\pm}$ emission components above the local continuum, and the $\langle|B|\rangle$ uncertainty is propagated from the uncertainty in these centroid positions.

To validate the use of this relation at the field strengths present in LP\,705-64, the measured $\upsigma^{\pm}$ component positions are compared against theoretical Zeeman transition wavelengths computed from atomic data \citep{Schimeczek2014}, as illustrated in Figure~\ref{fig:specpol}.  The observed emission features are overlaid onto a grid of theoretical Zeeman transition wavelengths as a function of field strength to visualise where the observed features intersect the theoretical curves.  This is carried out for each spectrum, and the field strength estimates obtained in this manner are in good agreement with those from Equation~\ref{eq:zeeman}, confirming that the classical relation remains applicable.
 
\begin{figure}
    \includegraphics[width=\columnwidth]{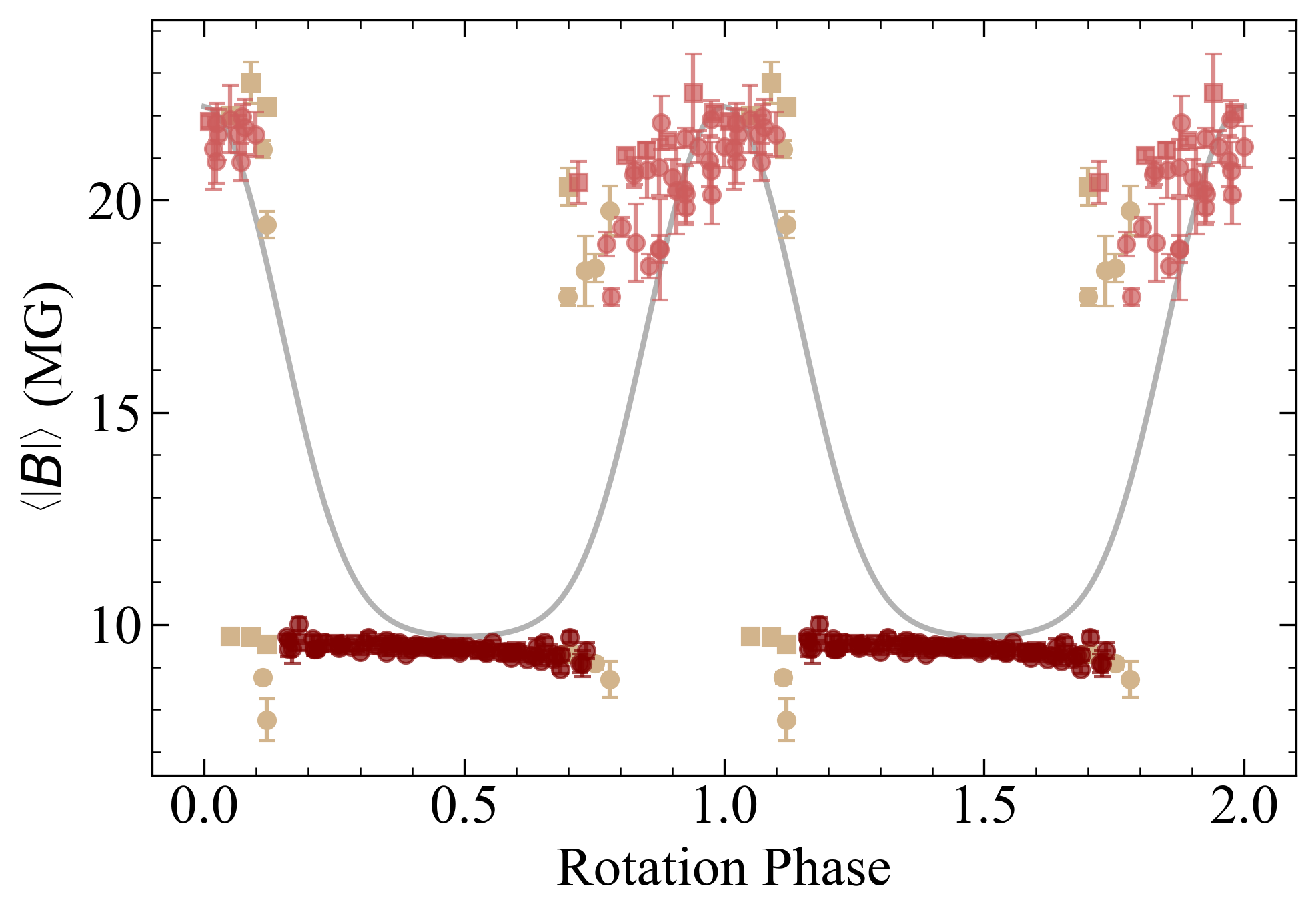}  
    \caption{$\langle |B| \rangle$ as a function of rotational phase, where each data point corresponds to a single exposure of either X-shooter or FORS2, measured using the H$\upalpha$ $\upsigma^{\pm}$ component splitting. The colour-coding is identical to that in Figures~\ref{fig:lcvsew} and \ref{fig:centralwavelengths}; however, owing to the inclusion of measurements using the FORS2 intensity spectra, there are additional data points coloured in tan, signifying those exposures where the timing resulted in the presence of both the narrow and broad Zeeman features.  In grey is a model for an offset dipole that results in the correct higher and lower field strengths as a function of rotation phase.  This particular model has a dipole centre offset by 0.27\,$R_\star$ toward the stronger pole, with a rotation axis inclined to the observer by $70\degr$, and a magnetic axis that is $70\degr$ from the axis of rotation.  Other parameter choices can reproduce the field strength extrema, but this example demonstrates that the observations are a departure from a relatively simple field configuration (see Section~4.2 for further details).}
    \label{fig:bfield}
\end{figure}

The resulting $\langle|B|\rangle$ values are plotted in Figure~\ref{fig:bfield} as a function of rotation phase, and include all measurements from both X-shooter and FORS2 spectra.  Six of the 118 X-shooter exposures and four of the 24 FORS2 spectra contain both the narrow and broad emission triplets, where the different fractions are a direct result of the exposure time differences.  In these spectra, each triplet is measured independently at those phases, resulting in two sets of $\langle|B|\rangle$ measurements.

The time-series $\langle|B|\rangle$ measurements reveal two key features.  First, the transition between the narrow and broad emission regimes is abrupt, where not a single spectrum shows any signature of a field strength intermediate to those plotted near phases 0.0 and 0.5 in Figure~\ref{fig:bfield}.  Second, the two regions display distinct, phase-dependent behaviour: the narrow line splitting remains approximately constant at 9.5\,MG throughout its visibility between phases 0.15 and 0.75, while the broad splitting increases modestly from around 20 to 22\,MG between phases 0.75 and 0.15. Together, these characteristics confirm that the two emission regions are physically distinct, compact, and localised on the stellar surface.

Using the computed wavelength positions of all $\upsigma$ components from both the X-shooter and FORS2 intensity spectra, the mean separations of the H$\upalpha$ $\upsigma^{\pm}$ components are $39.8\,\text{\AA}\,\text{MG}^{-1}$ and $35.8\,\text{\AA}\,\text{MG}^{-1}$ for the narrow and broad features, respectively.  The former is in excellent, and the latter in reasonable, agreement with the 40\,$\text{\AA}\,\text{MG}^{-1}$ linear conversion for weak fields \citep{Schimeczek2014}.

\subsection{Broadening of the Zeeman features}
\label{sec:3.3}

The observed Zeeman effect in stellar spectral lines is the result of all magnetic field contributions integrated over the visible hemisphere, weighted by the corresponding solid angle.  For example, if the magnetic field were completely uniform across the emitting region in LP\,705-64, then at sufficient spectral resolution, the $\upsigma$ components would consist of a discrete set of transitions at well-defined wavelengths, where the line profiles would reflect only the intrinsic spread of those specific atomic energy levels.  Any excess broadening beyond this theoretical value should therefore indicate additional physical mechanisms, in particular a gradient in the magnetic field strength across the emission region.  

To quantify this, the theoretical dispersion of the $\upsigma$ component wavelengths is computed at 9.5\,MG based on atomic data \citep{Schimeczek2014}, using the five narrow $\upsigma^{\pm}$ components, weighted by their respective dipole strengths.  This yields full widths at half maximum of 10.7 and 12.0\,\AA\ for the $\upsigma^{-}$ and $\upsigma^{+}$ components, respectively.  To measure the observed line widths, Gaussian profiles are fitted to the $\upsigma^{\pm}$ components in each individual exposure, using the telluric-corrected, continuum-normalized, and 3\,\AA-binned X-shooter spectra.  Taking the median full width at half maximum for all spectra when the narrow emission lines are present ($0.15 <$ phase $<0.75$), the observed values are 30.3 and 32.7\,\AA\ for the $\upsigma^{-}$ and $\upsigma^{+}$ components, respectively.  Both observed widths exceed the theoretical values, suggesting that the observed broadening of each feature is due to a spread in magnetic field strengths over each emission area.  This is not surprising, as even a simple dipole configuration would broaden the emission lines at any viewing geometry.

For the broad Zeeman splitting at 22\,MG, the linear scaling of individual H$\upalpha$ component wavelengths with field strength breaks down (Section~\ref{sec:3.2}), and a direct quantitative comparison between theoretical and observed $\upsigma$ component widths is therefore not possible.  However, the strongly blended and broadened profiles of the broad $\upsigma$ components are themselves qualitatively consistent with a magnetic field rapidly varying its field strength across the emitting region, smearing its components.  Therefore a common picture emerges for both the 9.5 and 22\,MG features, in which the magnetic field strength is not uniform across either emitting area, but varies across the surface of each chromospheric spot.

\subsection{A simple model for the size of a magnetic region}
\label{sec:3.4}

Using the phase-folded time-series $\langle|B|\rangle$ data plotted in Figure~\ref{fig:bfield}, two approximations are made for the phase interval during which the magnetic field fully switches over between 9.5 and 22\,MG, and thus one or the other region comes fully in and out of view on the stellar surface.  

First, the data are treated heuristically, and modelled with a simulated light curve with a transiting, limb-darkening-free planet using {\sc batman} \citep{Kreidberg2015}.\footnote{This approach is adopted because starspot tools such as {\sc lcurve} and {\sc starry} are unable to reproduce the sharp transitions.}  In this case, the transit depth is arbitrary and not used to constrain the size or geometry of the magnetic region, but simply to determine the full ingress and egress durations.  These are symmetric in the model and thus equal, and taken as the interval between where the curve exceeds 1 and 99\,per cent of the distance between the lower and higher magnetic field strengths, which are treated as constant and thus flat in the simulation.  This method yields an ingress (egress) duration of 0.043 in phase.  While this simulation is formally heuristic, it should represent the size of the region along the direction of rotation approximately, if not exactly.  Second, consecutive pairs of X-shooter spectra are searched for those which exhibit the narrow emission features, but where they are absent in antecedent or subsequent exposures.  From these, the shortest phase duration between any pair is 0.048.  It is important to note that this can be faster (as in the simulation), but the exposure durations limit the time resolution of this exercise.  

Using these approximations, the angular radius of at least one region can be no larger than $10\degr$.  Moreover, this size can be smaller if the transition between magnetic regions is faster than sampled by the X-shooter observing cadence; there are no spectra where any intermediate field strengths are observed, indicating a transition more rapid than a single exposure (for further discussion see Section~\ref{sec:4.2}).  Thus one or both of the magnetic regions are significantly smaller than the surface spots that give rise to the light curve dimming, which have been modelled to have $\uptheta\geq20\degr$ \citep{Farihi2023}.  However, there are also caveats with the photometric spot modelling because of simplifying assumptions such as circular geometry for two antipodal spots of equal size.  Nevertheless, at least one of the magnetic regions may be substantially more confined than its corresponding dark surface spot.

\section{Discussion}
\label{sec:4}

This section discusses the findings of all measurements obtained to date, including inferences that can be made about the size and geometry of the surface inhomogeneities, magnetic field, and chromospheric emission regions of LP\,705-64.  

\subsection{Implications of the rotationally modulated emission lines}

The $\langle W_{\uplambda} \rangle$ values vary in antiphase with the broadband photometry, with semi-amplitudes of 35 and 4.5\,per cent for the emission and the photometry, respectively.  This behaviour was first observed in full-phase coverage photometry and spectroscopy of the prototype DAHe white dwarf, GD\,356 \citep{Walters2021}.  Although the phase coverage for WD\,J1653$-$1001 is sparse, it appears that the emission line strength and photometric brightness are also anticorrelated in that DAHe star \citep{Elms2026}.  In both GD\,356 and WD\,J1653$-$1001 there is only a single region observed along the line of sight with a dark surface spot, a magnetic field, and Balmer line emission, but in LP\,705-64, the antiphase behaviour is seen in duplicate and for both regions associated with these phenomena.  Both DAHe stars WD\,J1616+5410 and SDSS\,J1252$-$0234 have double-peaked light curves similar to LP\,705-64 \citep{Manser2023, Reding2023}, but lack phase-resolved equivalent width measurements at present.  Thus, to date, the DAHe class is consistent with having dark surface spots associated with polar regions that are highly magnetised, above which sit optically thin heated regions; that is, chromospheres \citep{Walters2021}.

The antiphase behaviour indicates the temperature inversions are strongest near the geometric centre of the dark surface regions, and it is tempting to co-identify these regions with the magnetic poles.  Assuming the chromosphere is optically thin, the most natural means to change the emission line strength is for heated gas to pass into or out of view with rotation.  Based on Figure~\ref{fig:lcvsew}, this implies that the regions where the surface is relatively dark and those with a temperature inversion are well correlated in visibility, and thus likely in {\em spatial} extent.  However, based on the estimates of the size of at least one of the magnetic regions, the field appears more complex than e.g.\ an offset dipole that yields two unequal-strength poles at opposite ends of the magnetic axis.  Nevertheless, all measurements to date are consistent with intrinsic atmospheric heating that occurs primarily near the magnetic poles, where modelling has suggested this can be the result of internally driven currents that dissipate in the upper atmosphere \citep{Lanza2024}.

\subsection{Interpretation of the magnetic fields and abrupt transition}
\label{sec:4.2}

The phase-resolved spectroscopy and spectropolarimetry presented here are consistent with the previous field strength estimates based on sparse phase coverage and comparing Zeeman triplet wavelength shifts with atomic data \citep{Reding2023}.  With complete phase coverage and a cadence of approximately 3.6\,min with X-shooter, further details emerge.  The narrow Zeeman splitting corresponds to an approximately constant field strength of 9.5\,MG at all phases where it is present, whereas the broad splitting increases modestly from around 20 to 22\,MG while visible.  The narrow splitting appears for approximately 60\,per cent of the rotation cycle, and thus simple geometry indicates the 9.5\,MG pole lies closer to the rotational pole facing the observer.

However, the abruptness of the transitions between field strength regimes is inconsistent with a simple offset (decentred) dipole picture.  Figure~\ref{fig:bfield} and the analysis in Section~\ref{sec:3.4} indicate that one or both of these regions require little time to fully (dis)appear.  In an offset dipole, the polar field strengths can be distinct, but geometrically each magnetic cap will extend from the pole to the magnetic equator, and thus be large in angular extent (e.g.\ a dipole offset by $R_\star/2$ will have one polar region with angular radius $60\degr$).  Thus the time-series $\langle |B| \rangle$ would be expected to have a relatively gradual (sinusoidal) transition as any extended region comes into and out of view, as demonstrated by the simulation plotted in Figure~\ref{fig:bfield}.  Even fields more complex than an offset dipole are insufficient to account for the non-sinusoidal behaviour observed in both $\langle B_{\rm z} \rangle$ and $\langle |B| \rangle$. In particular, the most general approach is a nine-parameter magnetic field consisting of a dipole plus a planar quadrupole that is arbitrarily oriented \citep{Bagnulo1996}, but cannot account for the abrupt transition observed in LP\,705-64.  The data stand in stark contrast to these models, and the magnetic topology must be more complex than a combination of low-order fields.

The Zeeman splitting is the result of the visible hemisphere-integrated magnetic field, and thus via this metric it is not possible for the weaker 9.5\,MG region to supersede the stronger 22\,MG region when visible.  Therefore, the 22\,MG field region must be rotating in and out of view, and the nearly constant 9.5\,MG field region may rotate out of view or may remain visible at all phases; the data cannot differentiate between these two scenarios.  In Section~\ref{sec:3.4}, the shortest time between pairs of exposures (where one set of features either appears or disappears) is found to be 3.49\,min or 0.048 in phase.  However, because Zeeman splitting represents the integral of the visible hemisphere, it is not expected to simultaneously observe independent sets of Zeeman triplets from opposite magnetic poles in a dipolar configuration.  Therefore, the transition between these regions must be faster than a single exposure, which is close to 192\,s (3.2\,min) in X-shooter.  Furthermore, the transition is {\em sufficiently} fast that it does not smear: in the six spectra (of 118) where both sets of features are present, both $\upsigma^{\pm}$ pairs remain distinct.  This implies the transition is even more abrupt than estimated.

The FORS2 spectropolarimetric observations were designed to obtain high S/N Stokes $V/I$ profiles by co-adding exposures at matched rotational phases.  However, given the rapid rotation period of 72.63\,min, individual exposure times of 470--480\,s correspond to a non-negligible fraction of the spin cycle, and the resulting time averaging causes the circular polarization signal to be partially washed out within each exposure.  In four of the 24 exposures both sets of features appear; this reflects time-averaging across the transition rather than genuine co-existence.  Future observations with shorter individual exposure times would be required to resolve the Stokes $V$ profiles with sufficient phase resolution for a quantitative analysis.  Whether the stellar surface carries a global field beyond that observed in the emission regions, and if so what its configuration might be, are questions that cannot be answered with the present data alone.

The phase-resolved $\langle B_{\rm z} \rangle$ values place an additional constraint on the magnetic geometry.  Whereas $\langle|B|\rangle$ measures the total averaged field strength, $\langle B_{\rm z}\rangle$ samples only the line-of-sight component, so that their ratio, and its variation with rotation, depends on the orientation of the field relative to the observer \citep[e.g.][]{Stibbs1950, Preston1967, Landstreet1992}. The 22\,MG region is sampled near maximum visibility at phase 0.90 and has $|\langle B_{\rm z} \rangle/\langle |B| \rangle | \approx0.4$.  The 9.5\,MG region has $|\langle B_{\rm z} \rangle/\langle |B| \rangle | \approx0.5$ at the side-viewing phases 0.14 and 0.62, dropping to $\approx0.3$ at the central viewing phase 0.38.  The near-equality of $\langle B_{\rm z} \rangle$ at phases 0.14 and 0.62 is consistent with symmetric viewing of the same magnetic region. The smaller value at phase 0.38, when this region is most centrally presented, is expected if the field lines emerge approximately radially at the centre of the spot and splay outwards. Viewed from directly above, the splayed components partially cancel in projection along the line of sight, whereas at oblique angles the projected field is more nearly aligned with it.  This is qualitatively consistent with a localised field anchored at each spot.  

A comparison of LP\,705-64 with the prototype GD\,356 reveals a similar excess of observed line width over the theoretical uniform-field dispersion.  The prototype has a theoretical dispersion of around 7\,\AA\, compared with observed $\upsigma^{\pm}$ widths of 26 and 19\,\AA\ \citep{Walters2021}.  This is broadly consistent with the excess broadening found for LP\,705-64 at 9.5\,MG, and suggests that unresolved magnetic field gradients across chromospheric emission regions may be a common feature of DAHe stars.  GD\,356 exhibits a single varying field centred at $\langle|B|\rangle$ = 11.5\,MG, consistent with a single magnetic pole remaining in view throughout its rotation \citep{Walters2021}.  LP\,705-64 represents a more complex case in which both poles are observable.  GD\,356 almost certainly has a second magnetic pole permanently out of sight due to the viewing geometry, such that the underlying magnetic configuration across the class could be broadly similar, and the observed differences reflect inclination rather than a fundamental distinction in field structure.

\subsection{Intrinsic chromospheric activity}

As in previous work, multiple independent lines of evidence point to an intrinsic origin for the chromospheric emission.  First, and perhaps most compelling, is that all known DAHe stars cluster tightly in a compact region of the \textit{Gaia} HR diagram \citep{Ganiscke2020, Walters2021, Manser2023}.  With over 20 members of this class, the narrow range of masses and cooling ages is unlikely to arise by chance, and is a strong indication of a shared phase of stellar evolution.  Of specific interest, this clustering is consistent with a cooling halt due to the onset of $^{22}$Ne distillation \citep{Bedard2024}, or possibly the late-time emergence of fossil magnetism preserved in the core \citep{Isern2017,Blatman2024,Camisassa2024}.  Both possibilities suggest the DAHe white dwarf spectral morphology may be a magnetic field break-out phenomenon.  Binary evolution models can produce $^{22}$Ne enhancement in stellar mergers modestly more massive than 0.7\,M$_\odot$ \citep{Rui2026}. However, unlike the Q-branch \citep{Cheng2019}, no corresponding overdensity in the HR diagram has yet been predicted or observed for these objects.

Second, the time-series behaviours of LP\,705-64, SDSS\,J1252$-$0234, and WD\,J1616+5410 are each consistent with two distinct regions of correlated starspots, magnetic activity, and chromospheric emission.  Although only LP\,705-64 has data that can unambiguously demonstrate that all three phenomena are consistent with being co-located and sitting on two opposing poles, all data to date are consistent with heated regions that are $180\degr$ out of phase in rotation \citep{Manser2023,Farihi2023}.  An external source of atmospheric heating, such as a unipolar inductor or an interaction with an orbiting planet, would give rise to heated regions that are offset from the magnetic poles, and should be more similar to the footprints of Io and other moons of Jupiter \citep{Li1998, Kavanaugh2023}.  

Third, an external source of heating should pollute the white dwarf with heavy ions as in accreting binaries \citep{Ferrario1997}.  Barring hydrogen, to date there is no evidence for any other chemical species in DAHe white dwarfs.  In LP\,705-64, the X-shooter spectra cover the \ion{Ca}{ii} K line at 3934\,\AA, each exposure with S/N $>20$ (Figure~\ref{fig:trailedspec}), but only show features consistent with weak emission from the higher Balmer lines (Section~\ref{sec:3.1}).  In contrast, \ion{Ca}{ii} H \& K are two of the strongest emission features found in the solar chromosphere and in active low-mass stars. Weak \ion{Ca}{ii} H \& K emission has been detected in the absorption cores of two heavily polluted white dwarfs \citep{Klein2011,LeBourdais2025}, where it likely originates in the upper atmosphere (i.e.\ possibly a weak chromosphere from accretion heating).  The absence of Ca emission and absorption in LP\,705-64 is consistent with a pure hydrogen atmosphere \citep{Zuckerman2003, Koester2005}, and thus free of any heavy ions that would be expected in the unipolar inductor model \citep{Wickramasinghe2010,Ganiscke2020}.

There are two further barriers to external sources of atmospheric heating.  One is the lack of magnetospheric ions, whose presence is a prerequisite for any star-planet interaction model to operate.  The other is the rapid rotation often observed among DAHe stars, where several members rotate far too quickly to allow a coherent current loop between an orbiting body and the stellar surface \citep[see section~4.3 in][]{Walters2021}.

Currently, there is only a single model for intrinsic atmospheric heating in white dwarfs.  This model shows that $^{22}$Ne distillation is sufficiently vigorous to drive currents along field lines, which then heat the upper atmosphere \citep{Lanza2024}.  The model predicts atmospheric heating is maximal near the magnetic poles and absent at the equator, consistent with the localisation of chromospheric emission to two compact regions in LP\,705-64.  No coronal emission is expected in the intrinsic heating model, and thus X-ray observations of LP\,705-64 would provide an additional test of this prediction, beyond the non-detection in GD\,356 \citep{Weisskopf2007}.  Together with the clustering in mass and cooling ages, and the possibility that 0.7\,M$_\odot$ white dwarf mergers can result in enhanced $^{22}$Ne that in turn should lead to distillation \citep{Rui2026}, the intrinsic heating model is a promising framework for the DAHe phenomenon, but the precise mechanism powering DAHe chromospheres remains an open question.

\section{Conclusion}
\label{sec:5}

This paper reports phase-resolved time-series spectroscopy and spectropolarimetry of LP\,705-64.  The chromospheric emission strength and mean magnetic field vary with the previously established rotation period of 72.63\,min, with emission visible at all rotational phases.  The $\langle W_{\uplambda} \rangle$ of H$\upalpha$ varies in antiphase with the broadband photometry, with emission strongest when the white dwarf is dimmest, and demonstrates that the chromosphere and dark surface regions are correlated structures. Furthermore, the structure of the $\langle W_{\uplambda} \rangle$ time-series indicates the chromospheric regions have substantial sizes, comparable to the starspots causing the broad-band variations.  The photometric variability is dominated by changes in the stellar continuum, as the Balmer emission contributes negligibly to broadband flux, consistent with the findings for the prototype GD\,356 \citep{Walters2021}.

Two distinct chromospheric regions are observed, corresponding to a narrow and a broad set of Zeeman-split H$\upalpha$ emission triplets and consistent with $\langle |B| \rangle=9.5$ and 22\,MG, respectively.  The transitions between the two magnetic regions occur abruptly (within a few minutes), and are inconsistent with a relatively simple offset dipole.  This rapid transition constrains the 22\,MG emitting region to have an angular radius $\uptheta<10\degr$.  This is in contrast to the relatively large dark surface spots with which it appears to be co-located, which previous work has shown must be larger than $20\degr$ \citep{Farihi2023}. It appears the magnetic field is significantly more complex than an offset dipole, but further constraints are not possible without higher-cadence spectropolarimetry.

The chromospheric emission in LP\,705-64 is consistent with an intrinsic origin and the atmospheric heating model based on $^{22}$Ne distillation \citep{Lanza2024}.  However, several questions remain open. Extending full-phase coverage spectroscopy and spectropolarimetry to a larger sample would establish whether the two-spot geometry and opposite polarity magnetic regions seen here are typical, and would constrain whether the underlying field configurations are broadly similar across the class.  The origin and long-term evolution of the MG-level magnetic fields in DAHe objects remain ambiguous, and require a combination of time-resolved spectroscopy and spectropolarimetry, population studies, and theoretical development to resolve. 

Despite the clustering of DAHe objects in the HR diagram, there does not appear to be any associated overdensity.  Population synthesis modelling exploring mergers that enhance $^{22}$Ne across a range of remnant masses would therefore be useful.  However, in the case that magnetic fields decay with time, as potentially suggested by recent work on the younger and older kinematic members of the Q-branch \citep{OuldRouis2026}, it may be that the DAHe phenomenon is not as long-lived as any cooling delay.  The DAHe white dwarfs share the field strength properties of the broader magnetic population, yet reveal their fields in ways that ordinary magnetic white dwarfs do not, making them powerful probes into the origin and evolution of white dwarf magnetism.

\section*{Acknowledgements}

The authors acknowledge support from ESO staff members during this multi-instrument observation programme carried out in visitor mode at Paranal. JDL acknowledges past financial support from the Natural Sciences and Engineering Research Council of Canada, funding reference number 6377-2016.

\section*{Data Availability}
The data underlying this article are available in the ESO Science Archive Facility under ESO programme ID 110.23UW.002.


\bibliographystyle{mnras}
\bibliography{ref} 

@ARTICLE{Appenzeller1998,
       author = {{Appenzeller}, I. and {Fricke}, K. and {F{\"u}rtig}, W. and {G{\"a}ssler}, W. and {H{\"a}fner}, R. and {Harke}, R. and {Hess}, H.-J. and {Hummel}, W. and {J{\"u}rgens}, P. and {Kudritzki}, R.-P. and {Mantel}, K.-H. and {Meisl}, W. and {Muschielok}, B. and {Nicklas}, H. and {Rupprecht}, G. and {Seifert}, W. and {Stahl}, O. and {Szeifert}, T. and {Tarantik}, K.},
        title = "{Successful commissioning of FORS1 - the first optical instrument on the VLT.}",
      journal = {The Messenger},
         year = 1998,
        month = dec,
       volume = {94},
        pages = {1-6},
       adsurl = {https://ui.adsabs.harvard.edu/abs/1998Msngr..94....1A}
}

@ARTICLE{Bagnulo1996,
       author = {{Bagnulo}, S. and {Landi Degl'Innocenti}, M. and {Landi Degl'Innocenti}, E.},
        title = "{Multipolar magnetic fields in rotating AP stars: modeling of observable quantities.}",
      journal = {\aap},
         year = 1996,
        month = apr,
       volume = {308},
        pages = {115-131},
       adsurl = {https://ui.adsabs.harvard.edu/abs/1996A&A...308..115B}
}

@ARTICLE{Bagnulo2002,
       author = {{Bagnulo}, S. and {Szeifert}, T. and {Wade}, G.~A. and {Landstreet}, J.~D. and {Mathys}, G.},
        title = "{Measuring magnetic fields of early-type stars with FORS1 at the VLT}",
      journal = {\aap},
         year = 2002,
        month = jul,
       volume = {389},
        pages = {191-201},
          doi = {10.1051/0004-6361:20020606},
       adsurl = {https://ui.adsabs.harvard.edu/abs/2002A&A...389..191B}
}

@ARTICLE{Freudling2013,
       author = {{Freudling}, W. and {Romaniello}, M. and {Bramich}, D.~M. and {Ballester}, P. and {Forchi}, V. and {Garc{\'\i}a-Dabl{\'o}}, C.~E. and {Moehler}, S. and {Neeser}, M.~J.},
        title = "{Automated data reduction workflows for astronomy. The ESO Reflex environment}",
      journal = {\aap},
         year = 2013,
        month = nov,
       volume = {559},
          eid = {A96},
        pages = {A96},
          doi = {10.1051/0004-6361/201322494},
archivePrefix = {arXiv},
       eprint = {1311.5411},
 primaryClass = {astro-ph.IM},
       adsurl = {https://ui.adsabs.harvard.edu/abs/2013A&A...559A..96F}
}

@ARTICLE{Landstreet1975,
       author = {{Landstreet}, J.~D. and {Borra}, E.~F. and {Angel}, J.~R.~P. and {Illing}, R.~M.~E.},
        title = "{A search for strong magnetic fields in rapidly rotating Ap stars.}",
      journal = {\apj},
         year = 1975,
        month = nov,
       volume = {201},
        pages = {624-629},
          doi = {10.1086/153929},
       adsurl = {https://ui.adsabs.harvard.edu/abs/1975ApJ...201..624L}
}

@ARTICLE{Klein2011,
       author = {{Klein}, B. and {Jura}, M. and {Koester}, D. and {Zuckerman}, B.},
        title = "{Rocky Extrasolar Planetary Compositions Derived from Externally Polluted White Dwarfs}",
      journal = {\apj},
         year = 2011,
        month = nov,
       volume = {741},
       number = {1},
          eid = {64},
        pages = {64},
          doi = {10.1088/0004-637X/741/1/64},
archivePrefix = {arXiv},
       eprint = {1108.1565},
 primaryClass = {astro-ph.EP},
       adsurl = {https://ui.adsabs.harvard.edu/abs/2011ApJ...741...64K}
}

@ARTICLE{Bagnulo2015,
       author = {{Bagnulo}, S. and {Fossati}, L. and {Landstreet}, J.~D. and {Izzo}, C.},
        title = "{The FORS1 catalogue of stellar magnetic field measurements}",
      journal = {\aap},
         year = 2015,
        month = nov,
       volume = {583},
          eid = {A115},
        pages = {A115},
          doi = {10.1051/0004-6361/201526497},
archivePrefix = {arXiv},
       eprint = {1508.03839},
 primaryClass = {astro-ph.SR},
       adsurl = {https://ui.adsabs.harvard.edu/abs/2015A&A...583A.115B}
}

@ARTICLE{Gaia2023,
       author = {{Gaia Collaboration} and {Vallenari}, A. and {Brown}, A.~G.~A. and {Prusti}, T. and {de Bruijne}, J.~H.~J. and {Arenou}, F. and {Babusiaux}, C. and {Biermann}, M. and {Creevey}, O.~L. and {Ducourant}, C. and {Evans}, D.~W. and {Eyer}, L. and {Guerra}, R. and {Hutton}, A. and {Jordi}, C. and {Klioner}, S.~A. and {Lammers}, U.~L. and {Lindegren}, L. and {Luri}, X. and {Mignard}, F. and {Panem}, C. and {Pourbaix}, D. and {Randich}, S. and {Sartoretti}, P. and {Soubiran}, C. and {Tanga}, P. and {Walton}, N.~A. and {Bailer-Jones}, C.~A.~L. and {Bastian}, U. and {Drimmel}, R. and {Jansen}, F. and {Katz}, D. and {Lattanzi}, M.~G. and {van Leeuwen}, F. and {Bakker}, J. and {Cacciari}, C. and {Casta{\~n}eda}, J. and {De Angeli}, F. and {Fabricius}, C. and {Fouesneau}, M. and {Fr{\'e}mat}, Y. and {Galluccio}, L. and {Guerrier}, A. and {Heiter}, U. and {Masana}, E. and {Messineo}, R. and {Mowlavi}, N. and {Nicolas}, C. and {Nienartowicz}, K. and {Pailler}, F. and {Panuzzo}, P. and {Riclet}, F. and {Roux}, W. and {Seabroke}, G.~M. and {Sordo}, R. and {Th{\'e}venin}, F. and {Gracia-Abril}, G. and {Portell}, J. and {Teyssier}, D. and {Altmann}, M. and {Andrae}, R. and {Audard}, M. and {Bellas-Velidis}, I. and {Benson}, K. and {Berthier}, J. and {Blomme}, R. and {Burgess}, P.~W. and {Busonero}, D. and {Busso}, G. and {C{\'a}novas}, H. and {Carry}, B. and {Cellino}, A. and {Cheek}, N. and {Clementini}, G. and {Damerdji}, Y. and {Davidson}, M. and {de Teodoro}, P. and {Nu{\~n}ez Campos}, M. and {Delchambre}, L. and {Dell'Oro}, A. and {Esquej}, P. and {Fern{\'a}ndez-Hern{\'a}ndez}, J. and {Fraile}, E. and {Garabato}, D. and {Garc{\'\i}a-Lario}, P. and {Gosset}, E. and {Haigron}, R. and {Halbwachs}, J.-L. and {Hambly}, N.~C. and {Harrison}, D.~L. and {Hern{\'a}ndez}, J. and {Hestroffer}, D. and {Hodgkin}, S.~T. and {Holl}, B. and {Jan{\ss}en}, K. and {Jevardat de Fombelle}, G. and {Jordan}, S. and {Krone-Martins}, A. and {Lanzafame}, A.~C. and {L{\"o}ffler}, W. and {Marchal}, O. and {Marrese}, P.~M. and {Moitinho}, A. and {Muinonen}, K. and {Osborne}, P. and {Pancino}, E. and {Pauwels}, T. and {Recio-Blanco}, A. and {Reyl{\'e}}, C. and {Riello}, M. and {Rimoldini}, L. and {Roegiers}, T. and {Rybizki}, J. and {Sarro}, L.~M. and {Siopis}, C. and {Smith}, M. and {Sozzetti}, A. and {Utrilla}, E. and {van Leeuwen}, M. and {Abbas}, U. and {{\'A}brah{\'a}m}, P. and {Abreu Aramburu}, A. and {Aerts}, C. and {Aguado}, J.~J. and {Ajaj}, M. and {Aldea-Montero}, F. and {Altavilla}, G. and {{\'A}lvarez}, M.~A. and {Alves}, J. and {Anders}, F. and {Anderson}, R.~I. and {Anglada Varela}, E. and {Antoja}, T. and {Baines}, D. and {Baker}, S.~G. and {Balaguer-N{\'u}{\~n}ez}, L. and {Balbinot}, E. and {Balog}, Z. and {Barache}, C. and {Barbato}, D. and {Barros}, M. and {Barstow}, M.~A. and {Bartolom{\'e}}, S. and {Bassilana}, J.-L. and {Bauchet}, N. and {Becciani}, U. and {Bellazzini}, M. and {Berihuete}, A. and {Bernet}, M. and {Bertone}, S. and {Bianchi}, L. and {Binnenfeld}, A. and {Blanco-Cuaresma}, S. and {Blazere}, A. and {Boch}, T. and {Bombrun}, A. and {Bossini}, D. and {Bouquillon}, S. and {Bragaglia}, A. and {Bramante}, L. and {Breedt}, E. and {Bressan}, A. and {Brouillet}, N. and {Brugaletta}, E. and {Bucciarelli}, B. and {Burlacu}, A. and {Butkevich}, A.~G. and {Buzzi}, R. and {Caffau}, E. and {Cancelliere}, R. and {Cantat-Gaudin}, T. and {Carballo}, R. and {Carlucci}, T. and {Carnerero}, M.~I. and {Carrasco}, J.~M. and {Casamiquela}, L. and {Castellani}, M. and {Castro-Ginard}, A. and {Chaoul}, L. and {Charlot}, P. and {Chemin}, L. and {Chiaramida}, V. and {Chiavassa}, A. and {Chornay}, N. and {Comoretto}, G. and {Contursi}, G. and {Cooper}, W.~J. and {Cornez}, T. and {Cowell}, S. and {Crifo}, F. and {Cropper}, M. and {Crosta}, M. and {Crowley}, C. and {Dafonte}, C. and {Dapergolas}, A. and {David}, M. and {David}, P. and {de Laverny}, P. and {De Luise}, F. and {De March}, R.},
        title = "{Gaia Data Release 3. Summary of the content and survey properties}",
      journal = {\aap},
         year = 2023,
        month = jun,
       volume = {674},
          eid = {A1},
        pages = {A1},
          doi = {10.1051/0004-6361/202243940},
archivePrefix = {arXiv},
       eprint = {2208.00211},
 primaryClass = {astro-ph.GA},
       adsurl = {https://ui.adsabs.harvard.edu/abs/2023A&A...674A...1G}
}

@ARTICLE{Swan2026,
       author = {{Swan}, Andrew and {G{\"a}nsicke}, Boris T. and {Izquierdo}, Paula and {Koester}, Detlev and {Manser}, Christopher J. and {Rogers}, Laura K. and {Xu}, Siyi and {Aguilar}, J. and {Ahlen}, S. and {Prieto}, C. Allende and {Silva}, L. Beraldo e. and {Bianchi}, D. and {Brooks}, D. and {Castander}, F.~J. and {Claybaugh}, T. and {de la Macorra}, A. and {Dey}, A. and {Font-Ribera}, A. and {Forero-Romero}, J.~E. and {Gazta{\~n}aga}, E. and {Fusillo}, N. Gentile and {Gontcho}, S. Gontcho A. and {Gutierrez}, G. and {Guy}, J. and {Honscheid}, K. and {Inight}, K. and {Ishak}, M. and {Joyce}, R. and {Koposov}, S.~E. and {Landriau}, M. and {Le Guillou}, L. and {Li}, T.~S. and {Manera}, M. and {Meisner}, A. and {Miquel}, R. and {Moustakas}, J. and {Nadathur}, S. and {Najita}, J. and {P{\'e}rez-R{\`a}fols}, I. and {Percival}, W.~J. and {Prada}, F. and {Rossi}, G. and {Sanchez}, E. and {Schlegel}, D. and {Schubnell}, M. and {Seo}, H. and {Silber}, J. and {Sprayberry}, D. and {Tarl{\'e}}, G. and {Weaver}, B.~A. and {Zhou}, R. and {Zou}, H.},
        title = "{The DESI Data Release 1 white dwarf catalogue}",
      journal = {\mnras},
         year = 2026,
        month = sep,
          doi = {10.1093/mnras/stag1701},
archivePrefix = {arXiv},
       eprint = {2609.04314},
 primaryClass = {astro-ph.SR},
       adsurl = {https://ui.adsabs.harvard.edu/abs/2026MNRAS.tmp.1582S}
}

@ARTICLE{Kilic2026,
       author = {{Kilic}, Mukremin and {Bergeron}, Pierre and {Blouin}, Simon and {Moss}, Adam and {Green}, Matthew J. and {Jewett}, Gracyn and {Barrientos}, Manuel and {Albright}, Alexander L. and {Gleason}, Alexander R. and {Brown}, Warren R. and {Vernekar}, Nagaraj and {Hayden}, Michael R.},
        title = "{A Detailed Model Atmosphere Analysis of Cool White Dwarfs in DESI DR1}",
      journal = {\apj},
         year = 2026,
        month = aug,
       volume = {1006},
       number = {2},
          eid = {248},
        pages = {248},
          doi = {10.3847/1538-4357/ae825f},
archivePrefix = {arXiv},
       eprint = {2606.25377},
 primaryClass = {astro-ph.SR},
       adsurl = {https://ui.adsabs.harvard.edu/abs/2026ApJ..1006..248K}
}

@ARTICLE{Bedard2024,
       author = {{B{\'e}dard}, Antoine and {Blouin}, Simon and {Cheng}, Sihao},
        title = "{Buoyant crystals halt the cooling of white dwarf stars}",
      journal = {\nat},
         year = 2024,
        month = mar,
       volume = {627},
       number = {8003},
        pages = {286-288},
          doi = {10.1038/s41586-024-07102-y},
archivePrefix = {arXiv},
       eprint = {2409.04419},
 primaryClass = {astro-ph.SR},
       adsurl = {https://ui.adsabs.harvard.edu/abs/2024Natur.627..286B}
}

@ARTICLE{Preston1967,
       author = {{Preston}, George W.},
        title = "{A Statistical Investigation of the Orientation of Magnetic Axes in the Periodic Magnetic Variables}",
      journal = {\apj},
         year = 1967,
        month = nov,
       volume = {150},
        pages = {547},
          doi = {10.1086/149358},
       adsurl = {https://ui.adsabs.harvard.edu/abs/1967ApJ...150..547P}
}

@ARTICLE{Stibbs1950,
       author = {{Stibbs}, D.~W.~N.},
        title = "{A study of the spectrum and magnetic variable star HD 125248}",
      journal = {\mnras},
         year = 1950,
        month = jan,
       volume = {110},
        pages = {395},
          doi = {10.1093/mnras/110.4.395},
       adsurl = {https://ui.adsabs.harvard.edu/abs/1950MNRAS.110..395S}
}

@ARTICLE{Bagnulo2022,
       author = {{Bagnulo}, Stefano and {Landstreet}, John D.},
        title = "{Multiple Channels for the Onset of Magnetism in Isolated White Dwarfs}",
      journal = {\apjl},
         year = 2022,
        month = aug,
       volume = {935},
       number = {1},
          eid = {L12},
        pages = {L12},
          doi = {10.3847/2041-8213/ac84d3},
archivePrefix = {arXiv},
       eprint = {2208.02655},
 primaryClass = {astro-ph.SR},
       adsurl = {https://ui.adsabs.harvard.edu/abs/2022ApJ...935L..12B}
}

@ARTICLE{Koester2005,
       author = {{Koester}, D. and {Rollenhagen}, K. and {Napiwotzki}, R. and {Voss}, B. and {Christlieb}, N. and {Homeier}, D. and {Reimers}, D.},
        title = "{Metal traces in white dwarfs of the SPY (ESO Supernova Ia Progenitor Survey) sample}",
      journal = {\aap},
         year = 2005,
        month = mar,
       volume = {432},
       number = {3},
        pages = {1025-1032},
          doi = {10.1051/0004-6361:20041927},
       adsurl = {https://ui.adsabs.harvard.edu/abs/2005A&A...432.1025K}
}

@ARTICLE{Kavanaugh2023,
       author = {{Kavanagh}, Robert D. and {Vedantham}, Harish K.},
        title = "{Hunting for exoplanets via magnetic star-planet interactions: geometrical considerations for radio emission}",
      journal = {\mnras},
         year = 2023,
        month = oct,
       volume = {524},
       number = {4},
        pages = {6267-6284},
          doi = {10.1093/mnras/stad2035},
archivePrefix = {arXiv},
       eprint = {2307.02555},
 primaryClass = {astro-ph.EP},
       adsurl = {https://ui.adsabs.harvard.edu/abs/2023MNRAS.524.6267K}
}

@ARTICLE{Landstreet1992,
       author = {{Landstreet}, J.~D.},
        title = "{Magnetic fields at the surfaces of stars}",
      journal = {\aapr},
         year = 1992,
        month = jun,
       volume = {4},
       number = {1},
        pages = {35-77},
          doi = {10.1007/BF00873569},
       adsurl = {https://ui.adsabs.harvard.edu/abs/1992A&ARv...4...35L}
}

@ARTICLE{Elms2026,
       author = {{Elms}, Abbigail K. and {Bagnulo}, Stefano and {Tremblay}, Pier-Emmanuel and {Cunningham}, Tim and {Munday}, James and {Landstreet}, John and {El-Badry}, Kareem and {Caiazzo}, Ilaria and {Melis}, Carl and {Pinter}, Viktoria and {Weinberger}, Alycia},
        title = "{Detection of a weak magnetic field in the Balmer emission line white dwarf WDJ1653{\ensuremath{-}}1001}",
      journal = {\mnras},
         year = 2026,
        month = may,
       volume = {548},
       number = {1},
          eid = {stag505},
        pages = {stag505},
          doi = {10.1093/mnras/stag505},
archivePrefix = {arXiv},
       eprint = {2603.12048},
 primaryClass = {astro-ph.SR},
       adsurl = {https://ui.adsabs.harvard.edu/abs/2026MNRAS.548ag505E}
}

@ARTICLE{Kreidberg2015,
       author = {{Kreidberg}, Laura},
        title = "{batman: BAsic Transit Model cAlculatioN in Python}",
      journal = {\pasp},
         year = 2015,
        month = nov,
       volume = {127},
       number = {957},
        pages = {1161},
          doi = {10.1086/683602},
archivePrefix = {arXiv},
       eprint = {1507.08285},
 primaryClass = {astro-ph.EP},
       adsurl = {https://ui.adsabs.harvard.edu/abs/2015PASP..127.1161K}
}

@ARTICLE{Cheng2019,
       author = {{Cheng}, Sihao and {Cummings}, Jeffrey D. and {M{\'e}nard}, Brice},
        title = "{A Cooling Anomaly of High-mass White Dwarfs}",
      journal = {\apj},
         year = 2019,
        month = dec,
       volume = {886},
       number = {2},
          eid = {100},
        pages = {100},
          doi = {10.3847/1538-4357/ab4989},
archivePrefix = {arXiv},
       eprint = {1905.12710},
 primaryClass = {astro-ph.SR},
       adsurl = {https://ui.adsabs.harvard.edu/abs/2019ApJ...886..100C}
}

@ARTICLE{OuldRouis2026,
       author = {{Ould Rouis}, Lou Baya and {Hermes}, J.~J. and {Guidry}, Joseph A. and {Cheng}, Sihao and {Kilic}, Mukremin and {Vincent}, Olivier and {Bergeron}, Pierre and {Blouin}, Simon and {Moss}, Adam and {Lopez}, Isaac D. and {Jewett}, Gracyn},
        title = "{White Dwarf Merger Remnants with Cooling Delays on the Q Branch Lack Strong Magnetism}",
      journal = {arXiv e-prints},
         year = 2026,
        month = feb,
          eid = {arXiv:2602.02670},
        pages = {arXiv:2602.02670},
          doi = {10.48550/arXiv.2602.02670},
archivePrefix = {arXiv},
       eprint = {2602.02670},
 primaryClass = {astro-ph.SR},
       adsurl = {https://ui.adsabs.harvard.edu/abs/2026arXiv260202670O}
}

@ARTICLE{Bagnulo2019,
       author = {{Bagnulo}, S. and {Landstreet}, J.~D.},
        title = "{Discovery of weak magnetic fields in four DZ white dwarfs in the local 20 pc volume. Implications for the frequency of magnetic fields with cooling age}",
      journal = {\aap},
         year = 2019,
        month = oct,
       volume = {630},
          eid = {A65},
        pages = {A65},
          doi = {10.1051/0004-6361/201936068},
archivePrefix = {arXiv},
       eprint = {1908.08418},
 primaryClass = {astro-ph.SR},
       adsurl = {https://ui.adsabs.harvard.edu/abs/2019A&A...630A..65B}
}

@ARTICLE{Kawka2007,
       author = {{Kawka}, A. and {Vennes}, S. and {Schmidt}, G.~D. and {Wickramasinghe}, D.~T. and {Koch}, R.},
        title = "{Spectropolarimetric Survey of Hydrogen-rich White Dwarf Stars}",
      journal = {\apj},
         year = 2007,
        month = jan,
       volume = {654},
       number = {1},
        pages = {499-520},
          doi = {10.1086/509072},
archivePrefix = {arXiv},
       eprint = {astro-ph/0609273},
 primaryClass = {astro-ph},
       adsurl = {https://ui.adsabs.harvard.edu/abs/2007ApJ...654..499K}
}

@ARTICLE{Landstreet2012,
       author = {{Landstreet}, J.~D. and {Bagnulo}, S. and {Valyavin}, G.~G. and {Fossati}, L. and {Jordan}, S. and {Monin}, D. and {Wade}, G.~A.},
        title = "{On the incidence of weak magnetic fields in DA white dwarfs}",
      journal = {\aap},
         year = 2012,
        month = sep,
       volume = {545},
          eid = {A30},
        pages = {A30},
          doi = {10.1051/0004-6361/201219829},
archivePrefix = {arXiv},
       eprint = {1208.3650},
 primaryClass = {astro-ph.SR},
       adsurl = {https://ui.adsabs.harvard.edu/abs/2012A&A...545A..30L}
}

@ARTICLE{Goldoni2011,
       author = {{Goldoni}, P.},
        title = "{The X-shooter pipeline}",
      journal = {Astronomische Nachrichten},
         year = 2011,
        month = mar,
       volume = {332},
       number = {3},
        pages = {227},
          doi = {10.1002/asna.201111523},
       adsurl = {https://ui.adsabs.harvard.edu/abs/2011AN....332..227G}
}

@ARTICLE{Liebert1975,
       author = {{Liebert}, J. and {Angel}, J.~R.~P. and {Landstreet}, J.~D.},
        title = "{The detection of an Halpha Zeeman pattern in the cool magnetic white dwarf G 99-47.}",
      journal = {\apjl},
         year = 1975,
        month = dec,
       volume = {202},
        pages = {L139-L143},
          doi = {10.1086/181998},
       adsurl = {https://ui.adsabs.harvard.edu/abs/1975ApJ...202L.139L}
}

@ARTICLE{Kissin2015,
       author = {{Kissin}, Yevgeni and {Thompson}, Christopher},
        title = "{Spin and Magnetism of White Dwarfs}",
      journal = {\apj},
         year = 2015,
        month = aug,
       volume = {809},
       number = {2},
          eid = {108},
        pages = {108},
          doi = {10.1088/0004-637X/809/2/108},
archivePrefix = {arXiv},
       eprint = {1501.07197},
 primaryClass = {astro-ph.SR},
       adsurl = {https://ui.adsabs.harvard.edu/abs/2015ApJ...809..108K}
}

@ARTICLE{Wickramasinghe2005,
       author = {{Wickramasinghe}, D.~T. and {Ferrario}, Lilia},
        title = "{The origin of the magnetic fields in white dwarfs}",
      journal = {\mnras},
         year = 2005,
        month = feb,
       volume = {356},
       number = {4},
        pages = {1576-1582},
          doi = {10.1111/j.1365-2966.2004.08603.x},
       adsurl = {https://ui.adsabs.harvard.edu/abs/2005MNRAS.356.1576W}
}

@ARTICLE{Bagnulo2009,
       author = {{Bagnulo}, S. and {Landolfi}, M. and {Landstreet}, J.~D. and {Landi Degl'Innocenti}, E. and {Fossati}, L. and {Sterzik}, M.},
        title = "{Stellar Spectropolarimetry with Retarder Waveplate and Beam Splitter Devices}",
      journal = {\pasp},
         year = 2009,
        month = sep,
       volume = {121},
       number = {883},
        pages = {993},
          doi = {10.1086/605654},
       adsurl = {https://ui.adsabs.harvard.edu/abs/2009PASP..121..993B}
}

@ARTICLE{Reding2023,
       author = {{Reding}, Joshua S. and {Hermes}, J.~J. and {Clemens}, J.~C. and {Hegedus}, R.~J. and {Kaiser}, B.~C.},
        title = "{Two new white dwarfs with variable magnetic Balmer emission lines}",
      journal = {\mnras},
         year = 2023,
        month = jun,
       volume = {522},
       number = {1},
        pages = {693-699},
          doi = {10.1093/mnras/stad760},
archivePrefix = {arXiv},
       eprint = {2302.10207},
 primaryClass = {astro-ph.SR},
       adsurl = {https://ui.adsabs.harvard.edu/abs/2023MNRAS.522..693R}
}

@ARTICLE{Farihi2023,
       author = {{Farihi}, J. and {Hermes}, J.~J. and {Littlefair}, S.~P. and {Howarth}, I.~D. and {Walters}, N. and {Parsons}, S.~G.},
        title = "{Discovery of dipolar chromospheres in two white dwarfs}",
      journal = {\mnras},
         year = 2023,
        month = oct,
       volume = {525},
       number = {1},
        pages = {1097-1105},
          doi = {10.1093/mnras/stad2184},
archivePrefix = {arXiv},
       eprint = {2307.02543},
 primaryClass = {astro-ph.SR},
       adsurl = {https://ui.adsabs.harvard.edu/abs/2023MNRAS.525.1097F}
}

@ARTICLE{Schimeczek2014,
       author = {{Schimeczek}, C. and {Wunner}, G.},
        title = "{Atomic Data for the Spectral Analysis of Magnetic DA White Dwarfs in the SDSS}",
      journal = {\apjs},
         year = 2014,
        month = jun,
       volume = {212},
       number = {2},
          eid = {26},
        pages = {26},
          doi = {10.1088/0067-0049/212/2/26},
       adsurl = {https://ui.adsabs.harvard.edu/abs/2014ApJS..212...26S}
}

@ARTICLE{Walters2021,
       author = {{Walters}, N. and {Farihi}, J. and {Marsh}, T.~R. and {Bagnulo}, S. and {Landstreet}, J.~D. and {Hermes}, J.~J. and {Achilleos}, N. and {Wallach}, A. and {Hart}, M. and {Manser}, C.~J.},
        title = "{A test of the planet-star unipolar inductor for magnetic white dwarfs}",
      journal = {\mnras},
         year = 2021,
        month = may,
       volume = {503},
       number = {3},
        pages = {3743-3758},
          doi = {10.1093/mnras/stab617},
archivePrefix = {arXiv},
       eprint = {2103.01993},
 primaryClass = {astro-ph.SR},
       adsurl = {https://ui.adsabs.harvard.edu/abs/2021MNRAS.503.3743W}
}

@ARTICLE{Bagnulo2021,
       author = {{Bagnulo}, S. and {Landstreet}, J.~D.},
        title = "{New insight into the magnetism of degenerate stars from the analysis of a volume-limited sample of white dwarfs}",
      journal = {\mnras},
         year = 2021,
        month = nov,
       volume = {507},
       number = {4},
        pages = {5902-5951},
          doi = {10.1093/mnras/stab2046},
archivePrefix = {arXiv},
       eprint = {2106.11109},
 primaryClass = {astro-ph.SR},
       adsurl = {https://ui.adsabs.harvard.edu/abs/2021MNRAS.507.5902B}
}

@ARTICLE{Greenstein1985,
       author = {{Greenstein}, J.~L. and {McCarthy}, J.~K.},
        title = "{Emission lines in the magnetic white dwarf GD 356.}",
      journal = {\apj},
         year = 1985,
        month = feb,
       volume = {289},
        pages = {732-747},
          doi = {10.1086/162937},
       adsurl = {https://ui.adsabs.harvard.edu/abs/1985ApJ...289..732G}
}

@ARTICLE{Ganiscke2020,
       author = {{G{\"a}nsicke}, Boris T. and {Rodr{\'\i}guez-Gil}, Pablo and {Gentile Fusillo}, Nicola P. and {Inight}, Keith and {Schreiber}, Matthias R. and {Pala}, Anna F. and {Tremblay}, Pier-Emmanuel},
        title = "{Single magnetic white dwarfs with Balmer emission lines: a small class with consistent physical characteristics as possible signposts for close-in planetary companions}",
      journal = {\mnras},
         year = 2020,
        month = dec,
       volume = {499},
       number = {2},
        pages = {2564-2574},
          doi = {10.1093/mnras/staa2969},
archivePrefix = {arXiv},
       eprint = {2009.11925},
 primaryClass = {astro-ph.SR},
       adsurl = {https://ui.adsabs.harvard.edu/abs/2020MNRAS.499.2564G}
}

@ARTICLE{Manser2023,
       author = {{Manser}, Christopher J. and {G{\"a}nsicke}, Boris T. and {Inight}, Keith and {Robert}, Akshay and {Ahlen}, S. and {Allende Prieto}, C. and {Brooks}, D. and {Cooper}, A.~P. and {de la Macorra}, A. and {Font-Ribera}, A. and {Honscheid}, K. and {Kisner}, T. and {Landriau}, M. and {Meisner}, Aaron M. and {Miquel}, R. and {Nie}, Jundan and {Poppett}, C. and {Tarl{\'e}}, Gregory and {Zhou}, Zhimin},
        title = "{DAHe white dwarfs from the DESI Survey}",
      journal = {\mnras},
         year = 2023,
        month = jun,
       volume = {521},
       number = {4},
        pages = {4976-4994},
          doi = {10.1093/mnras/stad727},
archivePrefix = {arXiv},
       eprint = {2302.01335},
 primaryClass = {astro-ph.SR},
       adsurl = {https://ui.adsabs.harvard.edu/abs/2023MNRAS.521.4976M}
}

@ARTICLE{Reding2020,
       author = {{Reding}, Joshua S. and {Hermes}, J.~J. and {Vanderbosch}, Z. and {Dennihy}, E. and {Kaiser}, B.~C. and {Mace}, C.~B. and {Dunlap}, B.~H. and {Clemens}, J.~C.},
        title = "{An Isolated White Dwarf with 317 s Rotation and Magnetic Emission}",
      journal = {\apj},
         year = 2020,
        month = may,
       volume = {894},
       number = {1},
          eid = {19},
        pages = {19},
          doi = {10.3847/1538-4357/ab8239},
archivePrefix = {arXiv},
       eprint = {2003.10450},
 primaryClass = {astro-ph.SR},
       adsurl = {https://ui.adsabs.harvard.edu/abs/2020ApJ...894...19R}
}

@ARTICLE{Rui2026,
       author = {{Rui}, Nicholas Z. and {Fuller}, Jim},
        title = "{Forging Neon-distilling White Dwarfs in the Stellar Engulfments of Helium White Dwarfs}",
      journal = {\apj},
         year = 2026,
        month = feb,
       volume = {998},
       number = {1},
          eid = {64},
        pages = {64},
          doi = {10.3847/1538-4357/ae31e8},
archivePrefix = {arXiv},
       eprint = {2601.00072},
 primaryClass = {astro-ph.SR},
       adsurl = {https://ui.adsabs.harvard.edu/abs/2026ApJ...998...64R}
}

@ARTICLE{Lanza2024,
       author = {{Lanza}, A.~F. and {Rui}, N.~Z. and {Farihi}, J. and {Landstreet}, J.~D. and {Bagnulo}, S.},
        title = "{Atmospheric heating and magnetism driven by $^{22}$Ne distillation in isolated white dwarfs}",
      journal = {\aap},
         year = 2024,
        month = sep,
       volume = {689},
          eid = {A233},
        pages = {A233},
          doi = {10.1051/0004-6361/202449947},
archivePrefix = {arXiv},
       eprint = {2407.19289},
 primaryClass = {astro-ph.SR},
       adsurl = {https://ui.adsabs.harvard.edu/abs/2024A&A...689A.233L}
}

@ARTICLE{Lenz2005,
       author = {{Lenz}, P. and {Breger}, M.},
        title = "{Period04 User Guide}",
      journal = {Communications in Asteroseismology},
         year = "2005",
        month = "Jun",
       volume = {146},
        pages = {53-136},
          doi = {10.1553/cia146s53},
       adsurl = {https://ui.adsabs.harvard.edu/abs/2005CoAst.146...53L}
}

@ARTICLE{Blackett1947,
       author = {{Blackett}, P.~M.~S.},
        title = "{The Magnetic Field of Massive Rotating Bodies}",
      journal = {\nat},
         year = 1947,
        month = may,
       volume = {159},
       number = {4046},
        pages = {658-666},
          doi = {10.1038/159658a0},
       adsurl = {https://ui.adsabs.harvard.edu/abs/1947Natur.159..658B}
}

@ARTICLE{Kemp1970,
       author = {{Kemp}, James C. and {Swedlund}, John B. and {Landstreet}, J.~D. and {Angel}, J.~R.~P.},
        title = "{Discovery of Circularly Polarized Light from a White Dwarf}",
      journal = {\apjl},
         year = 1970,
        month = aug,
       volume = {161},
        pages = {L77},
          doi = {10.1086/180574},
       adsurl = {https://ui.adsabs.harvard.edu/abs/1970ApJ...161L..77K}
}

@ARTICLE{Camisassa2024,
       author = {{Camisassa}, M. and {Fuentes}, J.~R. and {Schreiber}, M.~R. and {Rebassa-Mansergas}, A. and {Torres}, S. and {Raddi}, R. and {Dominguez}, I.},
        title = "{Main sequence dynamo magnetic fields emerging in the white dwarf phase}",
      journal = {\aap},
         year = 2024,
        month = nov,
       volume = {691},
          eid = {L21},
        pages = {L21},
          doi = {10.1051/0004-6361/202452539},
archivePrefix = {arXiv},
       eprint = {2411.02296},
 primaryClass = {astro-ph.SR},
       adsurl = {https://ui.adsabs.harvard.edu/abs/2024A&A...691L..21C}
}

@ARTICLE{Isern2017,
       author = {{Isern}, Jordi and {Garc{\'\i}a-Berro}, Enrique and {K{\"u}lebi}, Baybars and {Lor{\'e}n-Aguilar}, Pablo},
        title = "{A Common Origin of Magnetism from Planets to White Dwarfs}",
      journal = {\apjl},
         year = 2017,
        month = feb,
       volume = {836},
       number = {2},
          eid = {L28},
        pages = {L28},
          doi = {10.3847/2041-8213/aa5eae},
archivePrefix = {arXiv},
       eprint = {1702.01813},
 primaryClass = {astro-ph.SR},
       adsurl = {https://ui.adsabs.harvard.edu/abs/2017ApJ...836L..28I}
}

@ARTICLE{Ginzburg2022,
       author = {{Ginzburg}, Sivan and {Fuller}, Jim and {Kawka}, Adela and {Caiazzo}, Ilaria},
        title = "{Slow convection and fast rotation in crystallization-driven white dwarf dynamos}",
      journal = {\mnras},
         year = 2022,
        month = aug,
       volume = {514},
       number = {3},
        pages = {4111-4119},
          doi = {10.1093/mnras/stac1363},
archivePrefix = {arXiv},
       eprint = {2202.12902},
 primaryClass = {astro-ph.SR},
       adsurl = {https://ui.adsabs.harvard.edu/abs/2022MNRAS.514.4111G}
}

@ARTICLE{Blatman2024,
       author = {{Blatman}, Daniel and {Ginzburg}, Sivan},
        title = "{Magnetic field breakout from white dwarf crystallization dynamos}",
      journal = {\mnras},
         year = 2024,
        month = feb,
       volume = {528},
       number = {2},
        pages = {3153-3162},
          doi = {10.1093/mnras/stae222},
archivePrefix = {arXiv},
       eprint = {2311.09299},
 primaryClass = {astro-ph.SR},
       adsurl = {https://ui.adsabs.harvard.edu/abs/2024MNRAS.528.3153B}
}

@ARTICLE{Garcia-Berro2012,
       author = {{Garc{\'\i}a-Berro}, Enrique and {Lor{\'e}n-Aguilar}, Pablo and {Aznar-Sigu{\'a}n}, Gabriela and {Torres}, Santiago and {Camacho}, Judit and {Althaus}, Leandro G. and {C{\'o}rsico}, Alejandro H. and {K{\"u}lebi}, Baybars and {Isern}, Jordi},
        title = "{Double Degenerate Mergers as Progenitors of High-field Magnetic White Dwarfs}",
      journal = {\apj},
         year = 2012,
        month = apr,
       volume = {749},
       number = {1},
          eid = {25},
        pages = {25},
          doi = {10.1088/0004-637X/749/1/25},
archivePrefix = {arXiv},
       eprint = {1202.0461},
 primaryClass = {astro-ph.SR},
       adsurl = {https://ui.adsabs.harvard.edu/abs/2012ApJ...749...25G}
}

@ARTICLE{Kilic2019,
       author = {{Kilic}, Mukremin and {Rolland}, B. and {Bergeron}, P. and {Vanderbosch}, Z. and {Benni}, P. and {Garlitz}, J.},
        title = "{A magnetic white dwarf with five H {\ensuremath{\alpha}} components}",
      journal = {\mnras},
         year = 2019,
        month = nov,
       volume = {489},
       number = {3},
        pages = {3648-3654},
          doi = {10.1093/mnras/stz2394},
archivePrefix = {arXiv},
       eprint = {1908.10915},
 primaryClass = {astro-ph.SR},
       adsurl = {https://ui.adsabs.harvard.edu/abs/2019MNRAS.489.3648K}
}

@ARTICLE{Vernet2011,
       author = {{Vernet}, J. and {Dekker}, H. and {D'Odorico}, S. and {Kaper}, L. and {Kjaergaard}, P. and {Hammer}, F. and {Randich}, S. and {Zerbi}, F. and {Groot}, P.~J. and {Hjorth}, J. and {Guinouard}, I. and {Navarro}, R. and {Adolfse}, T. and {Albers}, P.~W. and {Amans}, J. -P. and {Andersen}, J.~J. and {Andersen}, M.~I. and {Binetruy}, P. and {Bristow}, P. and {Castillo}, R. and {Chemla}, F. and {Christensen}, L. and {Conconi}, P. and {Conzelmann}, R. and {Dam}, J. and {de Caprio}, V. and {de Ugarte Postigo}, A. and {Delabre}, B. and {di Marcantonio}, P. and {Downing}, M. and {Elswijk}, E. and {Finger}, G. and {Fischer}, G. and {Flores}, H. and {Fran{\c{c}}ois}, P. and {Goldoni}, P. and {Guglielmi}, L. and {Haigron}, R. and {Hanenburg}, H. and {Hendriks}, I. and {Horrobin}, M. and {Horville}, D. and {Jessen}, N.~C. and {Kerber}, F. and {Kern}, L. and {Kiekebusch}, M. and {Kleszcz}, P. and {Klougart}, J. and {Kragt}, J. and {Larsen}, H.~H. and {Lizon}, J. -L. and {Lucuix}, C. and {Mainieri}, V. and {Manuputy}, R. and {Martayan}, C. and {Mason}, E. and {Mazzoleni}, R. and {Michaelsen}, N. and {Modigliani}, A. and {Moehler}, S. and {M{\o}ller}, P. and {Norup S{\o}rensen}, A. and {N{\o}rregaard}, P. and {P{\'e}roux}, C. and {Patat}, F. and {Pena}, E. and {Pragt}, J. and {Reinero}, C. and {Rigal}, F. and {Riva}, M. and {Roelfsema}, R. and {Royer}, F. and {Sacco}, G. and {Santin}, P. and {Schoenmaker}, T. and {Spano}, P. and {Sweers}, E. and {Ter Horst}, R. and {Tintori}, M. and {Tromp}, N. and {van Dael}, P. and {van der Vliet}, H. and {Venema}, L. and {Vidali}, M. and {Vinther}, J. and {Vola}, P. and {Winters}, R. and {Wistisen}, D. and {Wulterkens}, G. and {Zacchei}, A.},
        title = "{X-shooter, the new wide band intermediate resolution spectrograph at the ESO Very Large Telescope}",
      journal = {\aap},
         year = 2011,
        month = dec,
       volume = {536},
          eid = {A105},
        pages = {A105},
          doi = {10.1051/0004-6361/201117752},
archivePrefix = {arXiv},
       eprint = {1110.1944},
 primaryClass = {astro-ph.IM},
       adsurl = {https://ui.adsabs.harvard.edu/abs/2011A&A...536A.105V}
}

@ARTICLE{Angel1974,
       author = {{Angel}, J.~R.~P. and {Carswell}, R.~F. and {Strittmatter}, P.~A. and {Beaver}, E.~A. and {Harms}, R.},
        title = "{Discovery of a Magnetic DA White Dwarf}",
      journal = {\apjl},
         year = 1974,
        month = nov,
       volume = {194},
        pages = {L47},
          doi = {10.1086/181665},
       adsurl = {https://ui.adsabs.harvard.edu/abs/1974ApJ...194L..47A}
}

@ARTICLE{Schmidt2003,
       author = {{Schmidt}, Gary D. and {Harris}, Hugh C. and {Liebert}, James and {Eisenstein}, Daniel J. and {Anderson}, Scott F. and {Brinkmann}, J. and {Hall}, Patrick B. and {Harvanek}, Michael and {Hawley}, Suzanne and {Kleinman}, S.~J. and {Knapp}, Gillian R. and {Krzesinski}, Jurek and {Lamb}, Don Q. and {Long}, Dan and {Munn}, Jeffrey A. and {Neilsen}, Eric H. and {Newman}, Peter R. and {Nitta}, Atsuko and {Schlegel}, David J. and {Schneider}, Donald P. and {Silvestri}, Nicole M. and {Smith}, J. Allyn and {Snedden}, Stephanie A. and {Szkody}, Paula and {Vanden Berk}, Dan},
        title = "{Magnetic White Dwarfs from the Sloan Digital Sky Survey: The First Data Release}",
      journal = {\apj},
         year = 2003,
        month = oct,
       volume = {595},
       number = {2},
        pages = {1101-1113},
          doi = {10.1086/377476},
archivePrefix = {arXiv},
       eprint = {astro-ph/0307121},
 primaryClass = {astro-ph},
       adsurl = {https://ui.adsabs.harvard.edu/abs/2003ApJ...595.1101S}
}

@ARTICLE{Angel1970,
       author = {{Angel}, J.~R.~P. and {Landstreet}, J.~D.},
        title = "{Magnetic Observations of White Dwarfs}",
      journal = {\apjl},
         year = 1970,
        month = jun,
       volume = {160},
        pages = {L147},
          doi = {10.1086/180548},
       adsurl = {https://ui.adsabs.harvard.edu/abs/1970ApJ...160L.147A}
}

@ARTICLE{Jordan2007,
       author = {{Jordan}, S. and {Aznar Cuadrado}, R. and {Napiwotzki}, R. and {Schmid}, H.~M. and {Solanki}, S.~K.},
        title = "{The fraction of DA white dwarfs with kilo-Gauss magnetic fields}",
      journal = {\aap},
         year = 2007,
        month = feb,
       volume = {462},
       number = {3},
        pages = {1097-1101},
          doi = {10.1051/0004-6361:20066163},
archivePrefix = {arXiv},
       eprint = {astro-ph/0610875},
 primaryClass = {astro-ph},
       adsurl = {https://ui.adsabs.harvard.edu/abs/2007A&A...462.1097J}
}

@ARTICLE{Landstreet2019,
       author = {{Landstreet}, J.~D. and {Bagnulo}, S.},
        title = "{Discovery of kilogauss magnetic fields on the nearby white dwarfs WD 1105-340 and WD 2150+591}",
      journal = {\aap},
         year = 2019,
        month = mar,
       volume = {623},
          eid = {A46},
        pages = {A46},
          doi = {10.1051/0004-6361/201834638},
archivePrefix = {arXiv},
       eprint = {1901.10353},
 primaryClass = {astro-ph.SR},
       adsurl = {https://ui.adsabs.harvard.edu/abs/2019A&A...623A..46L}
}

@ARTICLE{Woltjer1964,
       author = {{Woltjer}, L.},
        title = "{X-Rays and Type I Supernova Remnants.}",
      journal = {\apj},
         year = 1964,
        month = oct,
       volume = {140},
        pages = {1309-1313},
          doi = {10.1086/148028},
       adsurl = {https://ui.adsabs.harvard.edu/abs/1964ApJ...140.1309W}
}

@ARTICLE{Landstreet1967,
       author = {{Landstreet}, John Darlington},
        title = "{Synchrotron Radiation of Neutrinos and Its Astrophysical Significance}",
      journal = {Physical Review},
         year = 1967,
        month = jan,
       volume = {153},
       number = {5},
        pages = {1372-1377},
          doi = {10.1103/PhysRev.153.1372},
       adsurl = {https://ui.adsabs.harvard.edu/abs/1967PhRv..153.1372L}
}

@ARTICLE{Tout2004,
       author = {{Tout}, Christopher A. and {Wickramasinghe}, Dayal T. and {Ferrario}, Lilia},
        title = "{Magnetic fields in white dwarfs and stellar evolution}",
      journal = {\mnras},
         year = 2004,
        month = dec,
       volume = {355},
       number = {3},
        pages = {L13-L16},
          doi = {10.1111/j.1365-2966.2004.08482.x},
       adsurl = {https://ui.adsabs.harvard.edu/abs/2004MNRAS.355L..13T}
}

@ARTICLE{Nordhaus2011,
       author = {{Nordhaus}, J. and {Wellons}, S. and {Spiegel}, D.~S. and {Metzger}, B.~D. and {Blackman}, E.~G.},
        title = "{Formation of high-field magnetic white dwarfs from common envelopes}",
      journal = {Proceedings of the National Academy of Science},
         year = 2011,
        month = feb,
       volume = {108},
       number = {8},
        pages = {3135-3140},
          doi = {10.1073/pnas.1015005108},
archivePrefix = {arXiv},
       eprint = {1010.1529},
 primaryClass = {astro-ph.SR},
       adsurl = {https://ui.adsabs.harvard.edu/abs/2011PNAS..108.3135N}
}

@ARTICLE{Siess1999,
       author = {{Siess}, Lionel and {Livio}, Mario},
        title = "{The accretion of brown dwarfs and planets by giant stars - II. Solar-mass stars on the red giant branch}",
      journal = {\mnras},
         year = 1999,
        month = oct,
       volume = {308},
       number = {4},
        pages = {1133-1149},
          doi = {10.1046/j.1365-8711.1999.02784.x},
archivePrefix = {arXiv},
       eprint = {astro-ph/9905235},
 primaryClass = {astro-ph},
       adsurl = {https://ui.adsabs.harvard.edu/abs/1999MNRAS.308.1133S}
}

@ARTICLE{Farihi2011,
       author = {{Farihi}, J. and {Dufour}, P. and {Napiwotzki}, R. and {Koester}, D.},
        title = "{The magnetic and metallic degenerate G77-50}",
      journal = {\mnras},
         year = 2011,
        month = jun,
       volume = {413},
       number = {4},
        pages = {2559-2569},
          doi = {10.1111/j.1365-2966.2011.18325.x},
archivePrefix = {arXiv},
       eprint = {1101.2203},
 primaryClass = {astro-ph.EP},
       adsurl = {https://ui.adsabs.harvard.edu/abs/2011MNRAS.413.2559F}
}

@ARTICLE{Horn1968,
       author = {{van Horn}, H.~M.},
        title = "{Crystallization of White Dwarfs}",
      journal = {\apj},
         year = 1968,
        month = jan,
       volume = {151},
        pages = {227},
          doi = {10.1086/149432},
       adsurl = {https://ui.adsabs.harvard.edu/abs/1968ApJ...151..227V}
}

@ARTICLE{Regos1995,
       author = {{Reg{\H{o}}s}, Enik{\H{o}} and {Tout}, Christopher A.},
        title = "{The effect of magnetic fields in common-envelope evolution on the formation of cataclysmic variables}",
      journal = {\mnras},
         year = 1995,
        month = mar,
       volume = {273},
       number = {1},
        pages = {146-156},
          doi = {10.1093/mnras/273.1.146},
       adsurl = {https://ui.adsabs.harvard.edu/abs/1995MNRAS.273..146R}
}

@ARTICLE{Tout2008,
       author = {{Tout}, C.~A. and {Wickramasinghe}, D.~T. and {Liebert}, J. and {Ferrario}, L. and {Pringle}, J.~E.},
        title = "{Binary star origin of high field magnetic white dwarfs}",
      journal = {\mnras},
         year = 2008,
        month = jun,
       volume = {387},
       number = {2},
        pages = {897-901},
          doi = {10.1111/j.1365-2966.2008.13291.x},
archivePrefix = {arXiv},
       eprint = {0805.0115},
 primaryClass = {astro-ph},
       adsurl = {https://ui.adsabs.harvard.edu/abs/2008MNRAS.387..897T}
}

@ARTICLE{Li1998,
       author = {{Li}, Jianke and {Ferrario}, Lilia and {Wickramasinghe}, Dayal},
        title = "{Planets around White Dwarfs}",
      journal = {\apjl},
         year = 1998,
        month = aug,
       volume = {503},
       number = {2},
        pages = {L151-L154},
          doi = {10.1086/311546},
       adsurl = {https://ui.adsabs.harvard.edu/abs/1998ApJ...503L.151L}
}

@ARTICLE{Wickramasinghe2010,
       author = {{Wickramasinghe}, Dayal T. and {Farihi}, Jay and {Tout}, Christopher A. and {Ferrario}, Lilia and {Stancliffe}, Richard J.},
        title = "{Does GD356 have a terrestrial planetary companion?}",
      journal = {\mnras},
         year = 2010,
        month = jun,
       volume = {404},
       number = {4},
        pages = {1984-1991},
          doi = {10.1111/j.1365-2966.2010.16417.x},
archivePrefix = {arXiv},
       eprint = {1002.1761},
 primaryClass = {astro-ph.SR},
       adsurl = {https://ui.adsabs.harvard.edu/abs/2010MNRAS.404.1984W}
}

@ARTICLE{Ferrario1997,
       author = {{Ferrario}, Lilia and {Wickramasinghe}, D.~T. and {Liebert}, J. and {Schmidt}, Gary D. and {Bieging}, John H.},
        title = "{The magnetic field and emission-line spectrum of the remarkable white dwarf GD356}",
      journal = {\mnras},
         year = 1997,
        month = jul,
       volume = {289},
       number = {1},
        pages = {105-116},
          doi = {10.1093/mnras/289.1.105},
       adsurl = {https://ui.adsabs.harvard.edu/abs/1997MNRAS.289..105F}
}

@ARTICLE{Zuckerman2003,
       author = {{Zuckerman}, B. and {Koester}, D. and {Reid}, I.~N. and {H{\"u}nsch}, M.},
        title = "{Metal Lines in DA White Dwarfs}",
      journal = {\apj},
         year = 2003,
        month = oct,
       volume = {596},
       number = {1},
        pages = {477-495},
          doi = {10.1086/377492},
       adsurl = {https://ui.adsabs.harvard.edu/abs/2003ApJ...596..477Z}
}

@ARTICLE{Weisskopf2007,
       author = {{Weisskopf}, Martin C. and {Wu}, Kinwah and {Trimble}, Virginia and {O'Dell}, Stephen L. and {Elsner}, Ronald F. and {Zavlin}, Vyacheslav E. and {Kouveliotou}, Chryssa},
        title = "{A Chandra Search for Coronal X-Rays from the Cool White Dwarf GD 356}",
      journal = {\apj},
         year = 2007,
        month = mar,
       volume = {657},
       number = {2},
        pages = {1026-1036},
          doi = {10.1086/510776},
archivePrefix = {arXiv},
       eprint = {astro-ph/0609585},
 primaryClass = {astro-ph},
       adsurl = {https://ui.adsabs.harvard.edu/abs/2007ApJ...657.1026W}
}

@ARTICLE{LeBourdais2025,
       author = {{Le Bourdais}, {\'E}rika and {Dufour}, Patrick and {Melis}, Carl and {Klein}, Beth L. and {Rogers}, Laura K. and {B{\'e}dard}, Antoine and {Debes}, John and {Messier}, Ashley and {Weinberger}, Alycia J. and {Xu}, Siyi},
        title = "{Tracing Planetary Accretion in a 3 Gyr old Hydrogen-rich White Dwarf: The Extremely Polluted Atmosphere of LSPM J0207+3331}",
      journal = {\apj},
         year = 2025,
        month = nov,
       volume = {993},
       number = {1},
          eid = {8},
        pages = {8},
          doi = {10.3847/1538-4357/ae0ace},
archivePrefix = {arXiv},
       eprint = {2510.08676},
 primaryClass = {astro-ph.SR},
       adsurl = {https://ui.adsabs.harvard.edu/abs/2025ApJ...993....8L}
}


\bsp    
\end{document}